\documentclass[twocolumn,english,prb,notitlepage,superscriptaddress,nobibnotes,nolongbibliography,floatfix]{revtex4-2}
\usepackage{graphicx}
\usepackage{amsmath}
\usepackage{amssymb}
\usepackage{bm}
\usepackage{enumerate}
\usepackage{xfrac}
\usepackage[caption=false]{subfig}
\usepackage[normalem]{ulem}
\usepackage[dvipsnames]{xcolor}
\usepackage{txfonts}
\usepackage[mathscr]{euscript}
\usepackage[pdftex,
		  pdflang={en-US},
            colorlinks=true,
            citecolor=blue,
            linkcolor=blue,
            urlcolor=blue,
            pdfauthor={},
	      pdftitle={CeCo2Ga8-neutron-Kondo}]{hyperref}
\usepackage{verbatim}
\usepackage{ulem}

\begin{document}

\title{Incommensurate spin fluctuations in one-dimensional Kondo metal CeCo$_2$Ga$_8$}

\author{Yixuan Huang}  
\affiliation{Theoretical Division and Center for Integrated Nanotechnologies, Los Alamos National Laboratory, Los Alamos, NM 87545, USA}
\affiliation{Computational Quantum Matter Research Team, RIKEN Center for Emergent Matter Science (CEMS), Saitama 351-0198, Japan}

\author{P. Murgatroyd}  
\affiliation{MPA-Q, Los Alamos National Laboratory, Los Alamos, NM 87545, USA}

\author{Yi Wu} 
\affiliation{Laboratory of Atomic and Solid State Physics, Department of Physics, Cornell University,
Ithaca, New York 14853, USA}

\author{M. Cook}  
\affiliation{MPA-Q, Los Alamos National Laboratory, Los Alamos, NM 87545, USA}

\author{P. F. S. Rosa}  
\affiliation{MPA-Q, Los Alamos National Laboratory, Los Alamos, NM 87545, USA}

\author{E. D. Bauer}  
\affiliation{MPA-Q, Los Alamos National Laboratory, Los Alamos, NM 87545, USA}

\author{Asish K. Kundu}  
\affiliation{National Synchrotron Light Source II, Brookhaven National Lab, Upton, New York 11973, USA}
\author{A. Rajapitamahuni}  
\affiliation{National Synchrotron Light Source II, Brookhaven National Lab, Upton, New York 11973, USA}
\author{E. Vescovo}  
\affiliation{National Synchrotron Light Source II, Brookhaven National Lab, Upton, New York 11973, USA}

\author{S. Zhang}  
\affiliation{Laboratory of Atomic and Solid State Physics, Department of Physics, Cornell University,
Ithaca, New York 14853, USA}
\author{V. Anil}  
\affiliation{Laboratory of Atomic and Solid State Physics, Department of Physics, Cornell University,
Ithaca, New York 14853, USA}

\author{P. Piyawongwatthana}  
\affiliation{Materials and Life Science Division, J-PARC Center, Japan Atomic Energy Agency, Tokai, Ibaraki 319-1195, Japan}
\affiliation{Institute for Solid State Physics, The University of Tokyo, Kashiwa, Chiba 277-8581, Japan}

\author{N. Murai}  
\affiliation{Materials and Life Science Division, J-PARC Center, Japan Atomic Energy Agency, Tokai, Ibaraki 319-1195, Japan}

\author{M. Kofu}  
\affiliation{Materials and Life Science Division, J-PARC Center, Japan Atomic Energy Agency, Tokai, Ibaraki 319-1195, Japan}
\affiliation{Institute for Solid State Physics, The University of Tokyo, Kashiwa, Chiba 277-8581, Japan} 

\author{K. M. Shen} 
\affiliation{Laboratory of Atomic and Solid State Physics, Department of Physics, Cornell University,
Ithaca, New York 14853, USA}
\affiliation{Kavli Institute at Cornell for Nanoscale Science, Cornell University, Ithaca, New York 14853, USA}

\author{F. Ronning} 
\affiliation{MPA-Q, Los Alamos National Laboratory, Los Alamos, NM 87545, USA}

\author{Jian-Xin Zhu} 
\affiliation{Theoretical Division and Center for Integrated Nanotechnologies, Los Alamos National Laboratory, Los Alamos, NM 87545, USA}

\author{A. Scheie}
\email{scheieao@gcc.edu} 
\affiliation{MPA-Q, Los Alamos National Laboratory, Los Alamos, NM 87545, USA}
\affiliation{Department of Physics, Grove City College, Grove City, PA 16127, USA}

\date{\today}


\begin{abstract}
We present an experimental and numerical study of the spin fluctuations in 1D Kondo metal CeCo$_2$Ga$_8$. Using inelastic neutron spectroscopy, we measure highly one-dimensional magnetism with low-energy incommensurate short-ranged magnetic fluctuations. ARPES similarly shows a highly one-dimensional electronic band structure, confirming the one-dimensional nature of the system. We use density matrix renormalization group (DMRG) simulations of the 1D Kondo lattice model to interpret the measured spectrum, which successfully reproduce the neutron scattering features. We are thus able to place CeCo$_2$Ga$_8$ within the emergent incommensurate phase of the 1D Kondo lattice phase diagram, and demonstrate that the Kondo lattice simulated non-perturbatively is an accurate microscopic model for heavy fermion physics. This shows CeCo$_2$Ga$_8$ to be one-dimensional despite its complexities, and reveals a coexistence of low-energy Kondo and magnetic features in its inelastic spectrum.  
\end{abstract}
\maketitle


\section{Introduction}

Heavy fermion materials, discovered nearly 50 years ago \cite{Andres_1975,Steglich_1979}, are a long standing puzzle in condensed matter physics \cite{Steward_1984_RMP,fisk1995physics,pavarini2015many,Wirth2016,Steglich_2016,shaginyan2022peculiar}. In these materials, interactions between a lattice of local moments and itinerant electrons produce ``strongly correlated'' phases, including quantum criticality \cite{si2010heavy}, non-Fermi liquids \cite{shaginyan2022peculiar}, and unconventional superconductivity  \cite{Steglich_2016}. Heavy fermion compounds seem to share a common phenomenology (summarized by Doniach's phase diagram \cite{Doniach1977}); but despite decades of effort, no microscopic model exists which can explain their behavior. 

Part of the difficulty is that strongly-interacting electron models are very challenging to simulate.     
Some heavy fermion systems have been successfully modeled \textit{ab-initio} in the limits of 
strong and weak coupling between local and itinerant spins
(c.f. CeIn$_3$ \cite{simeth2022microscopic} and CePd$_3$ \cite{goremychkin2018coherent} wherein simulations were able to match magnetic neutron spectra). 
However, these models were approximations, explicitly assuming either coherent magnon excitations simulated with spin wave theory (CeIn$_3$) or local fluctuations simulated with dynamical mean field theory (CePd$_3$), with neither very close to the intermediate quantum critical point where new phases of matter emerge. 
A much greater challenge is to describe a generic spectrum by a minimal theoretical model, where the collective dynamics are allowed to emerge from the calculations. 

One promising route is to study heavy fermions in one dimension. Here, the calculations are tractable using Density Matrix Renormalization Group (DMRG) theory \cite{schollwock2011density}, a nonperturbative low-dimensional numerical technique which can simulate the energy spectrum of a lattice model. 
In this way, we can directly test whether given models are realistic in describing a given system. 


\begin{figure}
	\centering
	\includegraphics[width=0.44\textwidth]{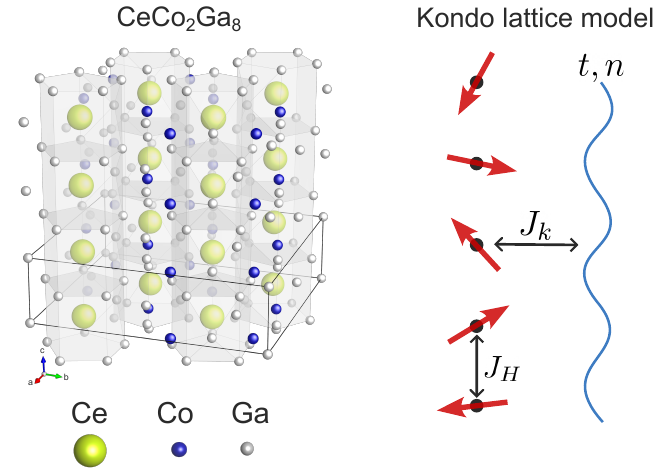}
	\caption{CeCo$_2$Ga$_8$ crystal structure (left) showing effective one-dimensional chains of Ce ions along the $c$-axis. We describe this with a simplified Kondo lattice model (right) where a one-dimensional chain of spins interact with each other with magnetic exchange interaction $J_H$ and also with an itinerant band (described by hopping $t$ and filling $n$) via Kondo interaction $J_k$.}
	\label{fig:schematic}
\end{figure}

In this study we focus on  CeCo$_2$Ga$_8$, a heavy fermion metal with one-dimensional chains of magnetic Ce ions extending along the $c$-axis of the orthorhombic unit cell, as shown in Fig. \ref{fig:schematic}. Electrical resistivity \cite{Cheng_2019_CeCo2Ga8}, density functional theory calculations \cite{wang2017heavy}, and polarized light spectroscopy \cite{Zheng_2022} all show a one-dimensional electronic state along the $c$ direction.  
No magnetic order has been observed down to 70~mK \cite{wang2017heavy,Bhattacharyya_2020}, and the zero-field specific heat diverges logarithmically while resistivity is linear with temperature \cite{wang2017heavy,Bhattacharyya_2020}, suggesting a quantum critical state. However, uniaxial strain appears to enhance the Sommerfeld coefficient, suggesting that ambient pressure CeCo$_2$Ga$_8$ is in the quantum critical region, but not quite at the quantum critical point \cite{Cheng_2022_CeCo2Ga8}. Regardless, this system (which also shows intriguing electronic Hall responses \cite{zou2024abnormal} and anomalous nuclear magnetic resonance signatures \cite{6d98-gndh}) appears to be a highly one-dimensional system in which heavy fermion physics can be found---an ideal system for comparing with theoretical calculations in one dimension. 

It is often assumed that the Kondo lattice model is the appropriate minimal model for Ce-based heavy fermions, and this model has been heavily studied in one dimension \cite{Tsunetsugu_1997} and where magnon quasiparticles are well-defined \cite{gao2024magnon}. However, the Kondo lattice model has not yet been compared to experimental neutron spectroscopy in the regime where quasiparticles break down. If a close agreement could be reached, this would be a significant validation of the Kondo lattice model as appropriate for real materials, and a breakthrough in interpreting the oft-inscrutable neutron spectra of heavy fermion materials. 

Two questions we aim to answer are: first, how one-dimensional the CeCo$_2$Ga$_8$ magnetic fluctuations are; and second, whether the system can be modeled by low-dimensional quantum simulation methods. We use neutron scattering and angle resolved photoemission spectroscopy (ARPES) to answer the first question, and DMRG to answer the second question. In the end, we show highly one-dimensional incommensurate behavior in CeCo$_2$Ga$_8$, and a qualitative match between the DMRG and CeCo$_2$Ga$_8$ which allows us to interpret the  CeCo$_2$Ga$_8$ spectral features. 
This shows firstly there is a coexistence of Kondo and magnetic features in the CeCo$_2$Ga$_8$ spectrum. Secondly, this shows that the single band Kondo lattice model accurately describes heavy fermion physics in 1D, suggesting that 2D and 3D systems can be (with sufficient computational power) be described by the same model. 

\section{Results}

\begin{figure*}
	\centering
	\includegraphics[width=0.99\textwidth]{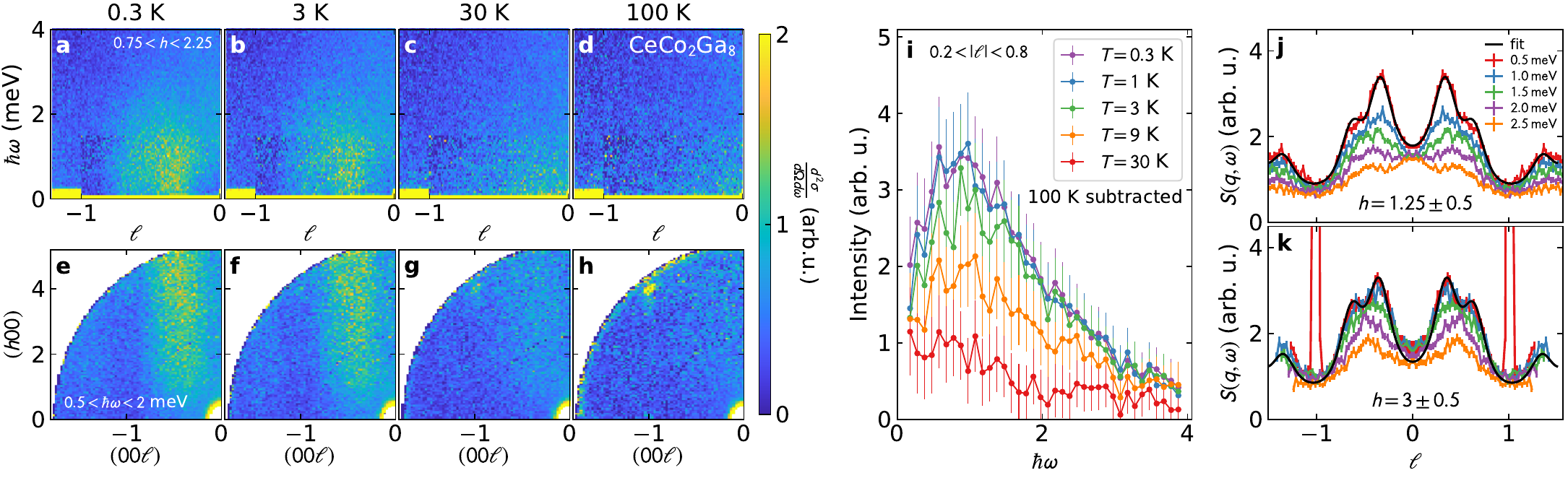}
	\caption{CeCo$_2$Ga$_8$ neutron scattering data. Panels {\bf a} through  {\bf d} show the energy-dependent scattering as a function of $\ell$ (along the chains). Panels {\bf e} through {\bf h} show the inelastic $0.5 <\hbar \omega < 2$ meV scattering along and transverse to the chains in reciprocal lattice units. Along the chains there is strong modulation, but transverse to the chains there is no momentum dependence. (Note that the $(h00)$ direction is actually an azimuthal average in the $(hk0)$ plane.)  Panel {\bf i} shows the energy-dependence of the $\ell=0.5$ intensity, showing a clear finite-energy maximum at $\approx 1$~meV. 
    Panels {\bf j} and {\bf k} show constant energy cuts of $T=0.3$~K CeCo$_2$Ga$_8$ neutron scattering data at $h=1.25$ and $h=3$ rlu. At low energies, there is a clear double-peak structure indicating incommensurate  $\ell = 0.354(2)$  fluctuations. At higher energies this coalesces to a single peak centered at $\ell = 0.5$. The black line indicates a double-peak fit with the polarization factor applied (see the supplemental information). Note the fit was performed only to panel \textbf{j}, and then recalculated for $h=3$ in panels \textbf{k}. 
    Note that panels {\bf a}-{\bf d} and {\bf i} show data composite of $E_i=5.9$~meV and $E_i=2.6$~meV, while panels {\bf e}-{\bf h} and {\bf j}-{\bf k} show $E_i=5.9$~meV data. 
    }
	\label{fig:neutrondata}
\end{figure*}

The inelastic CeCo$_2$Ga$_8$ neutron scattering is shown in Fig. \ref{fig:neutrondata}, the ARPES is shown in Fig. \ref{fig:ARPES}, and the density matrix renormalization group (DMRG) simulated spectra are shown in Fig. \ref{fig:DMRG}. Details of the experiments and calculations are given in the Methods section.

\subsection{Neutron Scattering}

At low temperatures, the CeCo$_2$Ga$_8$ spectrum shows a broad, diffuse continuum centered at $\ell=1/2$, signaling short-ranged antiferromagnetic (AFM) correlations along the chains (Fig. \ref{fig:neutrondata}{\bf a} - {\bf d}). 
Plotting the inelastic intensity along the chains versus transverse to the chains (Fig. \ref{fig:neutrondata}{\bf e} - {\bf h}) shows strong modulation along $\ell$ but almost no dependence on the $h$ direction (here $h$ is a superposition of all scattering vectors transverse to the chains because the coaligned crystals were only oriented along $\ell$), indicating a highly one-dimensional system. The one visible feature along $h$ is a suppression of intensity near $h=0$. This is explained by the polarization factor, wherein the neutron structure factor is only sensitive to spin correlations perpendicular to the scattering vector \cite{Squires}. Thus if CeCo$_2$Ga$_8$ spins fluctuate primarily along $c$ (as magnetic susceptibility shows \cite{wang2017heavy,Cheng_2019_CeCo2Ga8}), then this would suppress intensity where $\bf Q$ points along $\ell$---i.e., where $h=0$. This is precisely what the data show: polarization factor fits (shown in the supplemental information  \cite{SuppMat}) indicate over 90\% of the spin fluctuations are along $c$. 
Besides this low-$h$ suppression, the intensity transverse to the chains closely matches the isotropic Ce$^{3+}$ form factor \cite{BrownFF} (shown in the supplemental information \cite{SuppMat}), revealing a well-defined Ce$^{3+}$ effective $J=1/2$ doublet crystal field ground state with one-dimensional correlations.  

No dispersion is visible in the inelastic scattering (Fig. \ref{fig:neutrondata}{\bf a} - {\bf d}), so these excitations are not magnon-like. 
As shown in Fig. \ref{fig:neutrondata}{\bf i}, the intensity of the diffuse column is peaked at an energy of 1~meV, signaling that this system, although gapless and non-magnetically ordered, does have a characteristic energy scale which vanishes around $T^*=20$~K. As shown in the Supplemental Information, this feature precludes any quantum-critical scaling. 

At the lowest energies the peak intensity is found at an incommensurate wavevector along $\ell$, as shown in Fig. \ref{fig:neutrondata}{\bf j} - {\bf k}. The fitted incommensurate peak is found at $\ell = 0.354(2)$, with a fitted correlation length of 1.49(2) unit cells along the $c$ axis (see the Supplemental Materials for details \cite{SuppMat}). Including the polarization factor in the fit allows one to beautifully reproduce the observed scattering as shown in Fig. \ref{fig:neutrondata}{\bf j} - {\bf k} (see supplemental information for details \cite{SuppMat}). 

From the CeCo$_2$Ga$_8$ neutron scattering alone, we can conclude the system (i) is highly one-dimensional, (ii) has well-defined local Ce$^{3+}$ spins, (iii) has short-ranged correlations along the spin chains, (iv) the spins fluctuate primarily along the $c$-axis, (v) has excitations with a characteristic energy scale approximately at 1~meV, and (vi) has incommensurate fluctuations at low energies and commensurate antiferromagnetic fluctuations at high energies. 

\subsection{ARPES}

To investigate the dimensionality of the electronic structure in CeCo$_2$Ga$_8$, we used ARPES to map the momentum dependence of the states at the Fermi level. The effective dimensionality of an electronic state is reflected in the number of momentum directions over which it disperses appreciably. For a quasi-one-dimensional electronic structure, strong dispersion is expected along one principal momentum direction, while the transverse dispersion is strongly suppressed. In CeCo$_2$Ga$_8$, this requires weak variation along both $k_x$ and $k_y$ ($k_z$ is along the chain direction).

\begin{figure}
	\centering
	\includegraphics[width=\columnwidth]{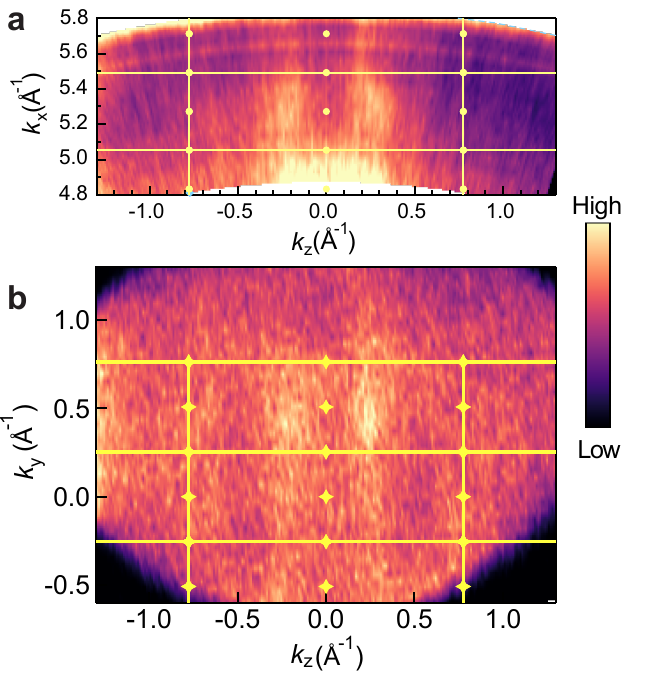}
	\caption{Highly anisotropic electronic structure of CeCo$_2$Ga$_8$. {\bf a} Out-of-plane ($k_x-k_z$) Fermi surface of CeCo$_2$Ga$_8$ measured over the photon energy range $h\nu =80-122$~eV. 
    {\bf b} In-plane ($k_x-k_y$) Fermi surface of CeCo$_2$Ga$_8$ measured at $h\nu = 115$~eV. Quasi-1D Fermi sheets are observed about the zone-center, consistent with a strongly suppressed transverse dispersion. The Brillouin zone is indicated in yellow. Fermi surfaces were obtained at $T = 18$~K and integrated over an energy window of $E_F \pm 15$~meV.
    }
	\label{fig:ARPES}
\end{figure}

Figure \ref{fig:ARPES} shows Fermi surface contours measured in the out-of-plane $k_z-k_x$ and in-plane $k_z-k_y$ momentum planes, respectively. In both measurements, Fermi sheets are observed near $k_z=\pm0.25$\AA$^{-1}$. These features show only weak modulation along both transverse momentum directions, indicating strongly reduced dispersion along $k_x$ and $k_y$. The observation of the same Fermi sheets at comparable $k_z$ positions in two orthogonal momentum planes provides direct evidence for a highly anisotropic, quasi-one-dimensional low-energy electronic structure in CeCo$_2$Ga$_8$.
We thus experimentally confirm CeCo$_2$Ga$_8$ as one-dimensional in both the magnetic and electronic channels (Fig. \ref{fig:schematic}). 
The question now is whether the observed behavior can be explained theoretically by the 1D Kondo lattice model.

\subsection{DMRG simulations}

\begin{figure*}
	\centering
	\includegraphics[width=\textwidth]{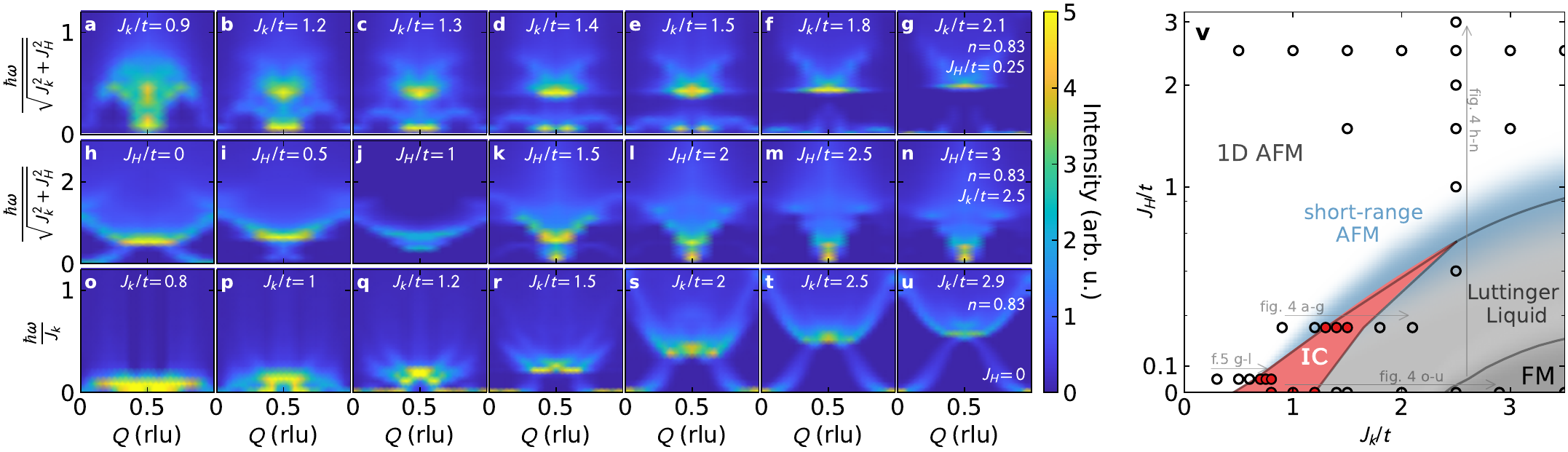}
	\caption{DMRG simulations of the 1D Kondo lattice model, calculating $S(Q,\omega)$. Panels on the left ({\bf a} - {\bf u}) show representative inelastic spectra tuning across the Luttinger Liquid phase boundary. Panel {\bf v} shows the $n=0.83$ simulation parameters on the phase diagram of Ref. \cite{Sikkema_1997}, but with the incommensurate (IC) phase added. Circles show the parameters we simulated, and the plotted spectra are referenced in light grey text. Points filled with red are where incommensuration is observed (\textit{e.g.}, panels {\bf c} - {\bf e}). In the 1D AFM phase (i.e., "spin gap" \cite{Sikkema_1997}) the spectrum resembles (and is continuously connected to) the 1D Heisenberg solution: a dispersive continuum with a lower boundary \cite{caux2006four}. However, on the boundary with the Luttinger Liquid phase the spectra broadens in momentum and, if $J_H$ is small enough, tunes through an incommensurate magnetic phase.}
	\label{fig:DMRG}
\end{figure*}

The DRMG simulations shown in Fig. \ref{fig:DMRG} show how the neutron spectrum $S(q,\omega)$ evolves from a 1D Heiesenberg-like spectrum (with the characteristic spinon continuum \cite{caux2006four}) to an antiferromagnetic singlet-triplet resonance. 
At large $J_H$ and small $J_K$, we observe antiferromagnetic correlations resembling the 1D spinon spectrum. Indeed, everywhere within the 1D AFM phase (mapped out with $n=0.83$ in previous studies \cite{Sikkema_1997,Tsunetsugu_1997,Peters_2012}) we observe peaked spectral weight at low energies and $Q=1/2$, indicating antiferromagnetism. (This phase is called ``spin gap'' in Ref. \cite{Sikkema_1997}, but because it is continuously connected to the gapless 1D Heisenberg chain we label it ``1D AFM''.)
Within the Luttinger Liquid (LL) state (identified theoretically by critical scaling of the Friedel oscillations \cite{Shibata_1996,Shibata_1997}), the system has a gapped short-ranged AFM fluctuation with a well-defined energy identified as the Kondo singlet-triplet excitation \cite{Ramasesha_1990}. Then, at small $J_H$ and large $J_k$, the low-energy spectral weight concentrates around $Q=0$ and $\hbar \omega = 0$ for a ferromagnetic (FM) state \cite{Peters_2012}. 

However, the most interesting behaviors emerge on the boundary between the LL and 1D AFM phase, Fig. \ref{fig:DMRG}({\bf a} - {\bf g}). As $J_k$ increases, the 1D Heisenberg dispersive spinon-like features are suppressed to lower energies, while the antiferromagnetic Kondo singlet-triplet resonance splits off and dominates the spectral weight at $J_k \gg J_H$. 
Meanwhile, near the boundary between the 1D AFM state and the LL state, the low-energy spectrum broadens in momentum and splits into two incommensurate peaks (c.f. Fig. \ref{fig:DMRG}{\bf d}-{\bf e}). 
This incommensurate phase is robust, appearing at various values of $J_H$ and $n$ even down to $J_H = 0$ (see below) and reflects competition between the AFM and kondo-stabilized FM state. 

\begin{figure*}
	\centering
	\includegraphics[width=0.85\textwidth]{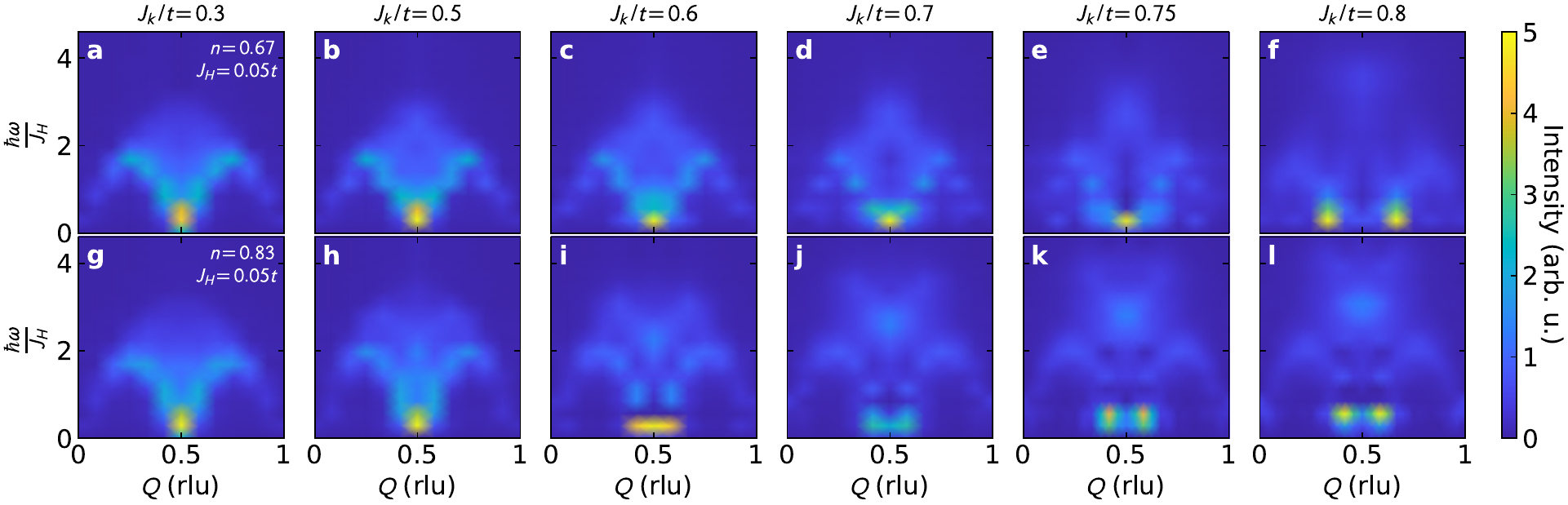}
	\caption{DMRG simulations of the 1D Kondo lattice model, with more realistic $J_H=0.05 t$ for CeCo$_2$Ga$_8$, at two different $n$ filling values. As in Fig. \ref{fig:DMRG}, the low-energy antiferromagnetic intensity concentrates and broadens as the system crosses the boundary into the Luttinger Liquid phase, with the $Q=0.5$ intensity splitting into two incommensurate wavevectors.}
	\label{fig:DMRG_realistic}
\end{figure*}

Helpful though they are, the simulation parameters used in Fig. \ref{fig:DMRG} are unrealistic for CeCo$_2$Ga$_8$. 
Firstly, we expect $J_H$ to be two or three orders of magnitude smaller than $t$ ($t = 0.56\pm0.07$~eV along $c$ ARPES fits \cite{SuppMat}, while $J_H \sim 0.6$~meV from susceptibility fits \cite{wang2017heavy}). So the realistic simulations will be in the lower left corner of the phase diagram in Fig. \ref{fig:DMRG}{\bf p}. 
Secondly, the appropriate filling factor $n$ for CeCo$_2$Ga$_8$ is unknown; we used $n=0.83$ in Fig. \ref{fig:DMRG} to compare with previous theoretical studies. 
To address the first point, it is computationally prohibitive to run DMRG simulations with $0 < J_H/t \lesssim 0.05$; but in Fig. \ref{fig:DMRG_realistic} we show calculations with $J_H/t = 0.05$ to get closer to the physical parameters. 
To address the second point, we run the simulations with different filling factors $n$, also shown in Fig. \ref{fig:DMRG_realistic}, to explore the AFM-LL phase boundary under different conditions.  

The results in Fig. \ref{fig:DMRG_realistic} show qualitatively similar behavior for both $n=0.83$ and $n=0.67$. As $J_k$ increases and the system crosses the 1D-AFM to LL phase boundary, the AFM resonance broadens in $Q$, loses its well-defined dispersion, and splits into incommensurate resonances at low energy. Meanwhile a weak $\ell=1/2$ triplet resonance appears at higher energy, which  splits off from the Heisenberg-like continuum as $J_k$ increases.  (In the supplemental information, we compute more values of $n$ and show that the spectrum in other ways does not strongly depend on $n$.) 

\section{Discussion}

The experimental results show that the CeCo$_2$Ga$_8$ is highly one-dimensional both magnetically and electronically with incommensurate magnetism at low energy. 
Despite the lack of magnetic order, the 1~meV maximum in intensity means this system (at least at ambient pressure) is not quantum critical.  
We note that this 1~meV energy scale is close to $T^* \sim 20$~K from resistivity \cite{Cheng_2019_CeCo2Ga8,wang2017heavy}, the temperature below which the system behaves as a coherent Kondo lattice (being below the expectation of incoherent electron scattering from multiple Kondo centers). 
Hence higher resolution data at lower energies is required to understand the non-Fermi liquid behavior found in transport and thermodynamic measurements. 

\begin{figure}
	\centering
	\includegraphics[width=0.9\columnwidth]{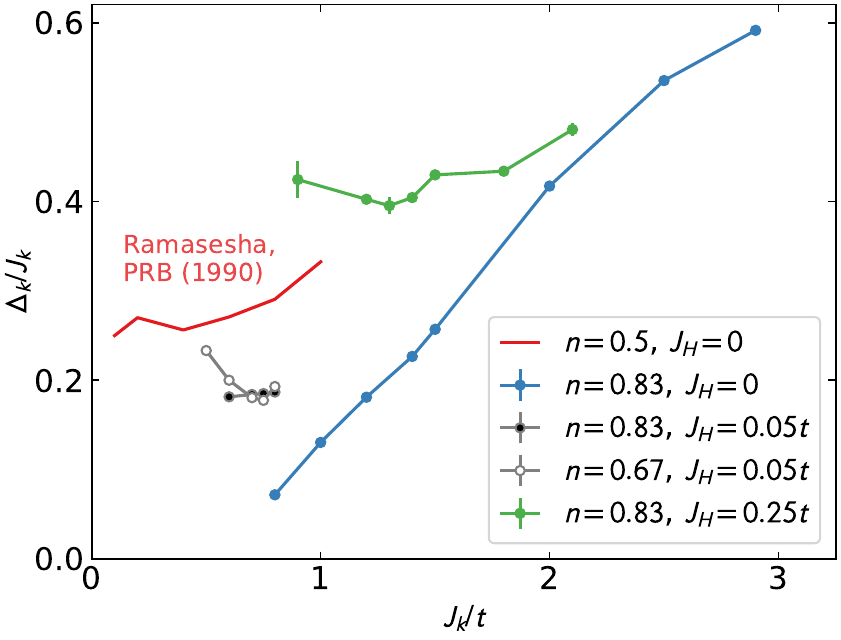}
	\caption{Evolution of the Kondo singlet-triplet resonance with $J_k$ from DMRG. Plotted points show the fitted resonances $\Delta$, alongside the $n=0.5$ result from Ref. \cite{Ramasesha_1990} in red, and the $J_H = 0.05 t$ simulations from Fig. \ref{fig:DMRG_realistic}.  The relationship between $J_k$ and $\Delta$ is in general nonlinear, and highly dependent upon $n$ and $J_H$.} 
	\label{fig:TripletExcitation}
\end{figure}

The DMRG results, meanwhile, show a fascinating non-perturbative view of the 1D  Kondo lattice model's magnetic excitations.  
In the limit of small $J_k$, one recovers the familiar 1D Heisenberg spectrum, but then the evolution with $J_K$ is highly nontrivial, involving intermediate phases depending on $n$, $J_k/t$, and $J_H/t$. 
The evolution of the Kondo singlet-triplet excitation is also nontrivial, depending strongly upon $J_H$ (as has been known for some time, c.f. Ref. \cite{Ramasesha_1990}). This is shown in Fig. \ref{fig:TripletExcitation}, which compares the evolution of the Kondo singlet-triplet resonance $\Delta_k$ of simulations with different parameters. For nonzero $J_H$, $\Delta_k \propto J_k$ for small $J_k$; whereas for $J_H=0$, $\Delta \propto J_k^2$ up to $J_k/t \sim 3$. 
This manifestly nonlinear behavior could (insofar as we may associate $\Delta_k$ with $T^*$) be associated with the flat bands  \cite{Kourris_2023} inherent in low-dimensional systems.


The existence of an incommensurate phase in the theoretical 1D Kondo Heisenberg model has been noted before \cite{Zachar_2001,Berg_2010,Tsvelik_2017,Tsvelik_2019}, and has been associated with charge-spin density waves of incipient unconventional superconductivity \cite{Zachar_2001,Berg_2010}, chiral spin liquids \cite{Tsvelik_2017} or helical metals \cite{Tsvelik_2019}. Our data do not allow us to distinguish a net spin chirality, but we note that no superconductivity has been found in CeCo$_2$Ga$_8$ down to 0.1~K \cite{wang2017heavy}. 

Of course, this Kondo-Heisenberg model neglects many details of CeCo$_2$Ga$_8$ (such as the presence of more than one itinerant band as revealed in DFT). 
Nevertheless, the 1D Kondo lattice model appears to be a reasonable minimal model. 
Although the match is imperfect, several of the observed features are found in the DMRG simulations. 
Firstly, the low-energy incommensuration is found near the boundary between the 1D AFM phase and the Luttinger Liquid. 
Secondly, the finite-energy $\ell = 1/2$ maximum at 1~meV can be associated with a Kondo singlet-triplet resonance. Such a resonance is theoretically found when the Kondo physics dominates the Heisenberg exchange ($J_k > J_H$). Third, a diffuse non-dispersive spectrum is also generically found in the incommensurate phase. 

An important difference between theory and experiment is that the CeCo$_2$Ga$_8$ spectrum is much broader in energy than any of the DMRG simulations. The experimental CeCo$_2$Ga$_8$ incommensurate fluctuations smoothly evolve into a commensurate triplet resonance at finite energy; whereas the DMRG tends to show  features more sharply separated in energy.  The reason for this discrepancy is unknown---perhaps higher dimensional electronic bands cause this broadening (Quantum Monte Carlo calculations of 1D Heisenberg chain Kondo coupled to 2D electronic bands show much more dramatic broadening in energy \cite{Liu_2023_Magnetic}). Or perhaps this broadened spectrum is the behavior of the Kondo-Heisenberg model in the limit where $0 < J_H < J_k \ll  t$, which our numerical calculations could not access. 


Nevertheless, 
the qualitative interpretation of the CeCo$_2$Ga$_8$ spectrum is clear: 
its low-energy incommensuration indicates a competition between 1D AFM and Luttinger Liquid behavior, and the finite energy maximum indicates an incipient Kondo singlet-triplet resonance. (This is strikingly similar to the Gapped dispersionless excitations near quantum criticality seen in other numerical approaches to the 1D Kondo lattice \cite{SciPostPhys.17.2.034}.) Because of the unknown filling factor $n$, we cannot assign precise values to $J_H$ and $J_k$. 
Still, our results strongly validate both CeCo$_2$Ga$_8$  as a 1D heavy fermion metal, and the 1D Kondo lattice as an appropriate model for simulating and interpreting it.  
This is an important step toward assigning rigorous meaning to the spectral features of heavy fermions, and thus toward solving one of the longest standing materials puzzles in physics.

\section{Summary and conclusion}

We have shown that  CeCo$_2$Ga$_8$ magnetic and electronic fluctuations are very one-dimensional, the magnetic excitations have a characteristic energy scale, and show incommensurate low-energy magnetic fluctuations along the spin chains. 
DMRG simulations of the 1D Kondo lattice show that these features are found on the boundary between the 1D AFM and Luttinger Liquid phase, when $J_H < J_k < t$. 
This means that CeCo$_2$Ga$_8$ is found within the incommensurate phase in Fig. \ref{fig:DMRG}. It also means that the Kondo exchange dominates the Heisenberg exchange, and $\ell = 1/2$ intensity maximum can be interpreted as an incipient singlet-triplet resonance. 

More broadly, this study shows how non-perturbative first-principles modeling can be used to interpret inelastic spin spectra of  heavy fermions. The CeCo$_2$Ga$_8$ magnetic spectrum is not sharp and spinon-like, where the spectrum can be modeled with long-ranged quasiparticles; nor is it completely short-ranged and diffuse, where the spectrum can be modeled as local fluctuations via dynamical mean field theory (DMFT). Instead its spectrum is between these limits, where both quasiparticle theory and DMFT fail. It is in this regime where  nontrivial incommensuration emerges, and in which a non-perturbative quantum simulation like DMRG is necessary to capture the behavior. 

This result is important not just for understanding CeCo$_2$Ga$_8$, but for understanding heavy fermion metals generally. The Kondo lattice model, though highly simplified, successfully reproduces the spectrum of a real physical material, revealing a superposition of incommensurate low-energy scattering and singlet-triplet high-energy scattering.  
We anticipate that the Kondo lattice model will continue to generate insight into heavy fermion materials: as numerical tools advance to allow 2D simulations, comparing numerics to low-dimensional materials is a workable strategy to finally solving this long-standing physics problem. 

\section{Methods}

\subsection{Neutron scattering}

We measured the inelastic neutron spectrum of CeCo$_2$Ga$_8$ using the AMATERAS spectrometer \cite{AMATERAS} at J-PARC. The sample consisted of 25 crystals totaling 2.6~g coaligned with the $c$-axis in the scattering plane (see supplemental information for details).
The sample was mounted in a $^3$He refrigerator and measured at 0.3~K, 1~K, 3~K, 9~K, 30~K, and 100~K for 20 hours at each temperature, rotating the sample 180$^{\circ}$ in one degree steps. Using rep-rate multiplication, we simultaneously collected data from $E_i = 23.6$~meV neutrons, $E_i = 5.9$~meV neutrons, and $E_i = 2.6$~meV neutrons, corresponding to elastic FWHM resolution 1.24~meV, 0.15~meV, and 0.064~meV respectively. 
The data are shown in Fig. \ref{fig:neutrondata}. 

\subsection{ARPES}

We performed angle-resolved photoemission spectroscopy (ARPES) measurements at the 21-ID ESM beamline of the National Synchrotron Light Source II (NSLS-II), Brookhaven National Laboratory, USA. 
Samples were cleaved in situ at $T=18$~K under ultra-high vacuum conditions with a pressure better than $2 \times 10^{-11}$ Torr. The overall energy resolution was better than 20~meV, while the angular resolution was $0.1^{\circ}$. Photon-energy-dependent measurements were acquired over the range $h\nu = 80–122$~eV. Out-of-plane momentum ($k_z$) values were determined using a free-electron final-state approximation with an inner potential of $V_0 = 15$~eV. All measurements were performed using linearly vertically polarized light. 
In this work the $k_z$ is defined as being along the $c$-axis, which is parallel to the one-dimensional Ce-chain. The orthogonal, transverse momentum directions are defined as $k_x$ and $k_y$. 

\subsection{DMRG simulations}

We simulated the inelastic neutron spectrum of the one-dimensional Kondo-Heisenberg model using DMRG. 
The Hamiltonian for the spin-1/2 Kondo-Heisenberg model on a one dimensional chain of length $L$ is given as 
\begin{equation}
\label{eq:H}
\mathcal{H}=-t{\displaystyle {\displaystyle {\textstyle {\displaystyle \sum_{i=1,\sigma}^{L-1}c_{i,\sigma}^{\dagger}}c_{i+1,\sigma}+h.c.}}+J_{K}\sum_{i=1}^{L}\boldsymbol{S}_{i}\cdot\boldsymbol{s}_{i}}+J_{H}\sum_{i=1}^{L-1}\boldsymbol{S}_{i}\cdot \boldsymbol{S}_{i+1}
\end{equation}
where $c_{i,\sigma}^{\dagger}$ refers to the electron creation operator on site $i$ with spin index $\sigma$; $\boldsymbol{S}_{i}$ refers to the localized spin-$\frac{1}{2}$ operator on site $i$; $\boldsymbol{s}_{i}=\frac{1}{2}\sum_{\alpha,\beta}c_{i,\alpha}^{\dagger}\boldsymbol{\sigma}_{\alpha,\beta}c_{i,\beta}$ refers to the conduction electron spin operator with the Pauli matrices $\boldsymbol{\sigma}_{\alpha,\beta}$. 
This model thus gives us four free parameters: hopping $t$, filling $n$, Kondo exchange $J_K$, and Heisenberg exchange $J_H$ (see Fig. \ref{fig:schematic}). 

The spectral function, also known as the dynamical spin structure factor, is calculated from the time-resolved localized spin-spin correlations in real space which is defined as
\begin{equation}
\label{eq:structure}
\chi^{SS}(q,\omega)=\int_{0}^{T}d\tau e^{i\omega\tau-\eta\tau}\frac{1}{L_{0}^{2}}\sum_{i,j\in L_{0}}e^{-iq(r_{j}-r_{i})}\left\langle \boldsymbol{S}_{i}(\tau) \cdot \boldsymbol{S}_{j}(0) \right\rangle
\end{equation}
The transformation to frequency space is done by a numerical integration with a smearing factor $\eta = 1/T$ to compromise the finite total time $T$ in the simulations. In order to avoid the open boundary effect we choose a segment of $L_{0}=L/4$ in the middle of the chain to perform the Fourier transformation.

\acknowledgments

We gratefully acknowledge the U.S. Department of Energy, Office of Basic Energy Sciences, Division of Materials Science and Engineering under project ``Quantum Fluctuations in Narrow-Band Systems.'' DMRG calculations were done with support from LANL LDRD Program and Center for Integrated Nanotechnologies, a DOE BES user facility, in partnership with the LANL Institutional Computing Program for computational resources. 
The neutron experiment at the Materials and Life Science Experimental Facility of the J-PARC was performed under a user program (Proposal No. 2023A0012). 
The ARPES results presented utilized resources at the ESM (21-ID-1) beamline of the National Synchrotron Light Source II, a U.S. Department of Energy (DOE) Office of Science User Facility operated by Brookhaven National Laboratory under Contract No. DE-SC0012704.
S.C. and V.A. acknowledge support from the National Science Foundation NSF DMR-2104427.


%

\pagebreak

\section*{Supplemental Information for Incommensurate spin fluctuations in one-dimensional Kondo metal $\bf{CeCo_2Ga_8}$} 

\quad 

\renewcommand{\thesection}{\Roman{section}}
\setcounter{section}{0}
\renewcommand\thefigure{S.\arabic{figure}}
\setcounter{figure}{0}
\renewcommand{\theequation}{S\arabic{equation}}
\setcounter{equation}{0}
\renewcommand{\thetable}{S\Roman{table}}
\setcounter{table}{0}


\section{Neutron experiment and analysis}

\subsection{Sample synthesis}

Single crystals of CeCo$_2$Ga$_8$ were synthesized using a Ga-In flux. The optimal starting ratio of constituent elements was found to be 1:1.25:12.75:2.25 = Ce:Co:Ga:In. The raw elements (Ce ingot, Co powder, Ga shot, and In chunk) were loaded into a 5 ml alumina crucible, then flame sealed in an evacuated quartz ampoule. The ampoule was then heated in a box furnace to $1100^{\circ}\textrm{C}$, homogenized at temperature for 10 hours, and subsequently cooled at a rate of $4^{\circ}\textrm{C}/\textrm{h}$ to $600^{\circ}\textrm{C}$. The crystals were then decanted from the melt with a centrifuge. The addition of indium into the melt resulted in large and separable rods of CeCo$_2$Ga$_8$. Attempts made using pure Ga self-flux were unable to produce the desired phase. 

\subsection{Neutron experiment details}

\begin{figure}
	\centering
	\includegraphics[width=0.4\textwidth]{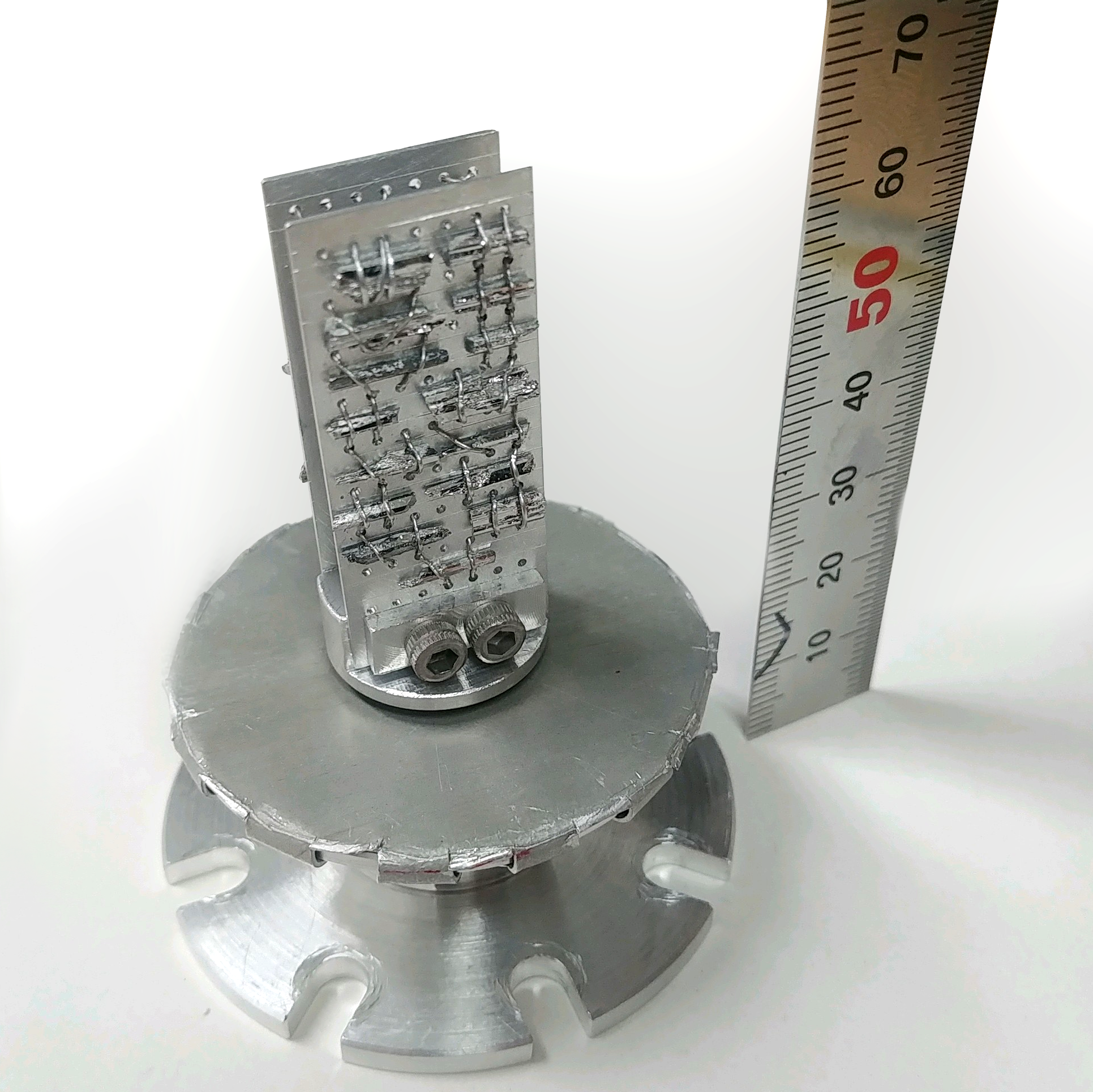}
	\caption{CeCo$_2$Ga$_8$ crystals measured in neutron spectroscopy, consisting of 2.6~g of coaligned crystals. The long axis of each crystal is the $c$ axis, which is aligned in the horizontal scattering plane. The azimuthal $ab$-plane angle is not aligned.}
	\label{fig:crystalmount}
\end{figure}

\begin{figure*}
	\centering
	\includegraphics[width=0.9\textwidth]{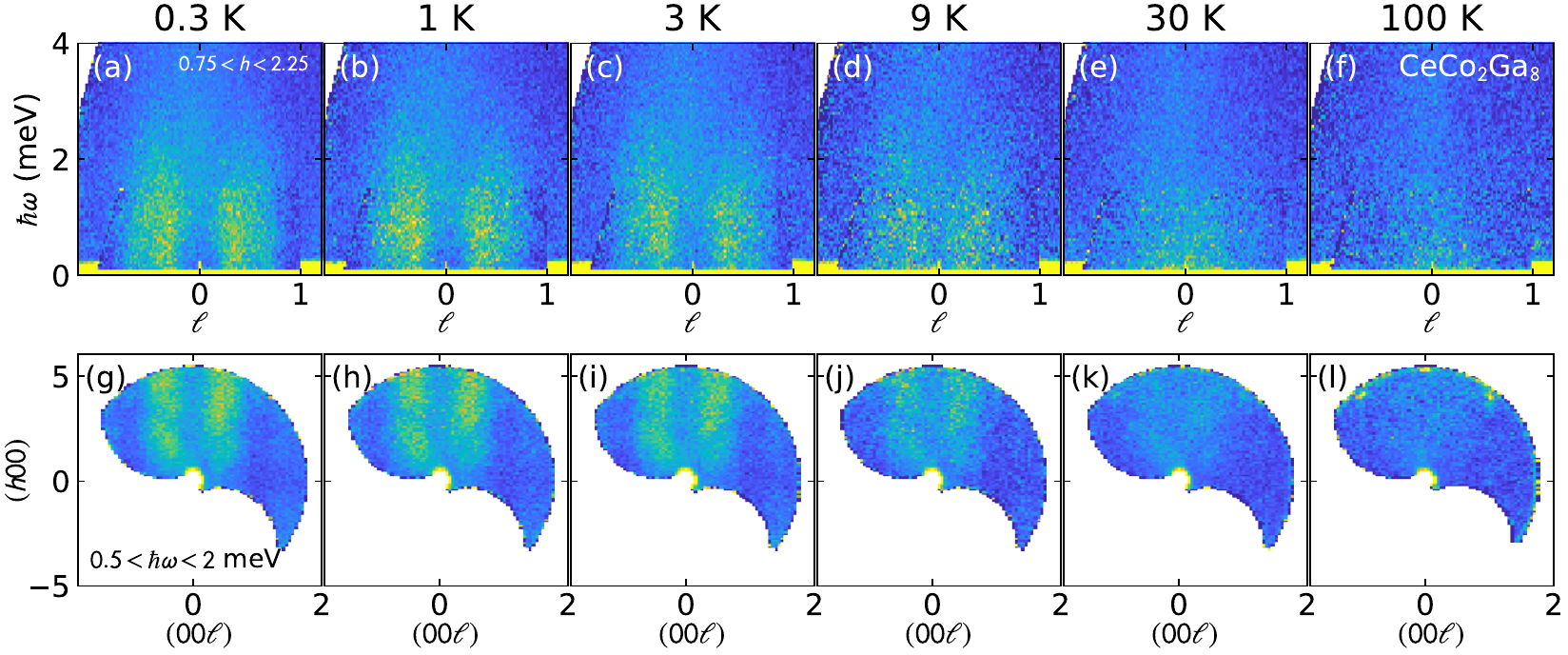}
	\caption{Unsymmetrized CeCo$_2$Ga$_8$ neutron scattering at different temperatures. The upper panels (a)-(f) show the energy-dependent scattering, and the lower panels (g)-(l) show constant-energy slices. Data shown in e.g. Fig. \ref{fig:HOL_comprehensive} have been folded about $\ell=0$ and $h=0$, but the overall trend is clear even without this symmetrization.}
	\label{fig:unsymmetrized}
\end{figure*}

\begin{figure*}
	\centering
	\includegraphics[width=0.8\textwidth]{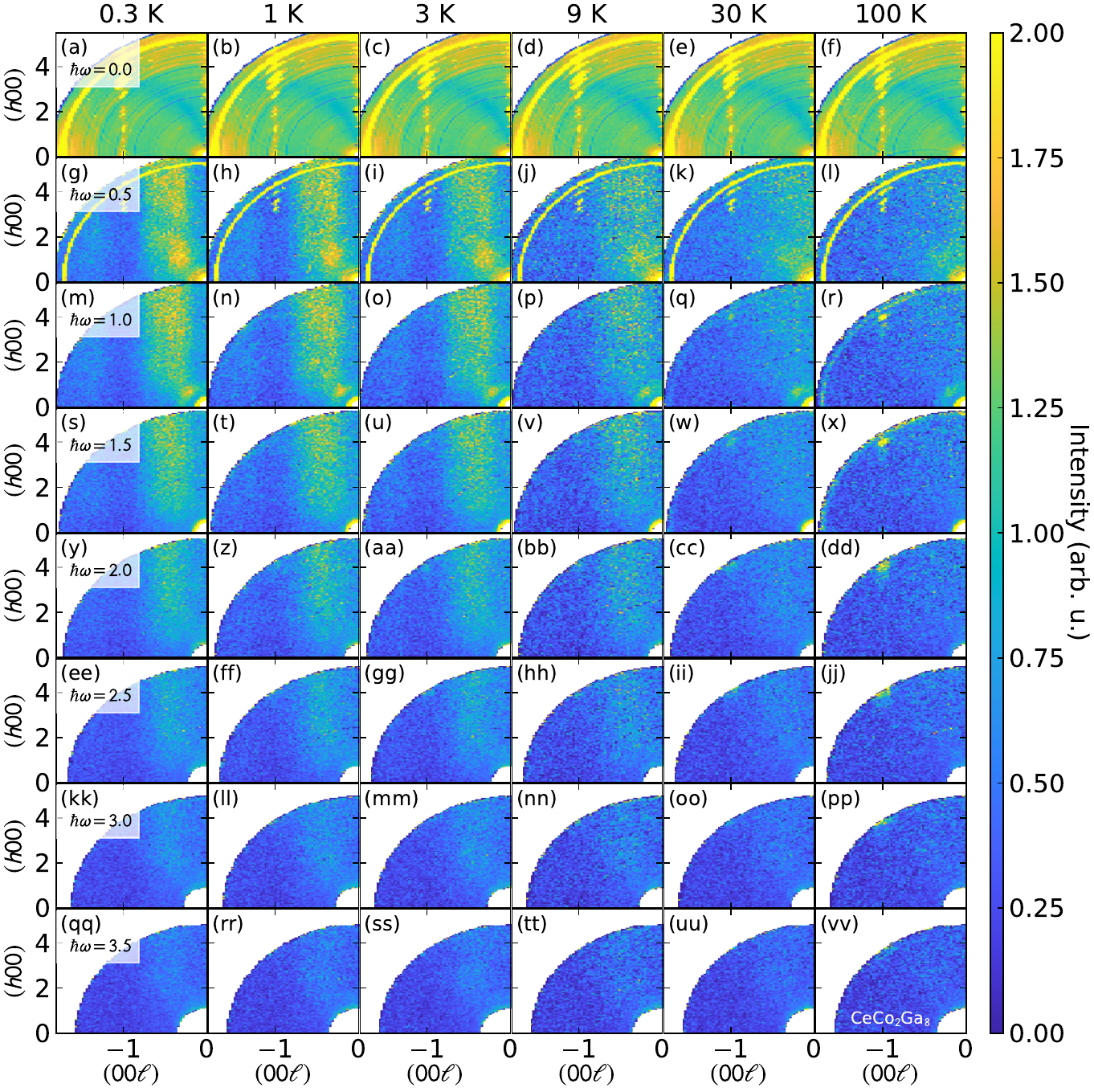}
	\caption{Constant-energy slices of CeCo$_2$Ga$_8$ neutron scattering as a function of temperature and energy transfer. Note the arc of reduced intensity in the low-energy scattering starting from (0,0,0) and moving out to (4,0,-1), which is an artifact of neutron absorption when the incident or final neutron is parallel to the aluminum plates.}
	\label{fig:HOL_comprehensive}
\end{figure*}

\begin{figure}
	\centering
	\includegraphics[width=0.5\textwidth]{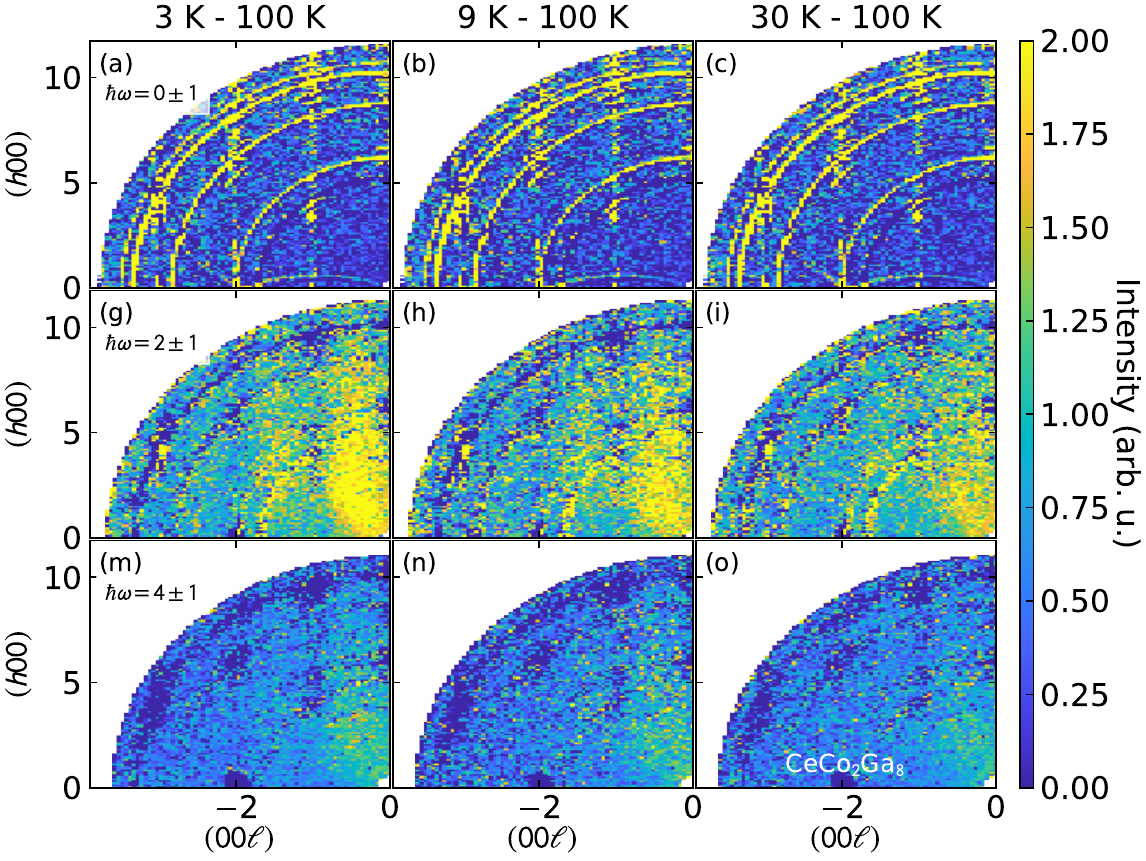}
	\caption{Temperature-subtracted CeCo$_2$Ga$_8$ scattering at  $E_i = 23.6$~meV, showing the evolution of the scattering pattern with energy transfer and temperature.}
	\label{fig:Ei23}
\end{figure}

\begin{figure}[h]
\vspace{5pt}
	\centering
	\includegraphics[width=0.42\textwidth]{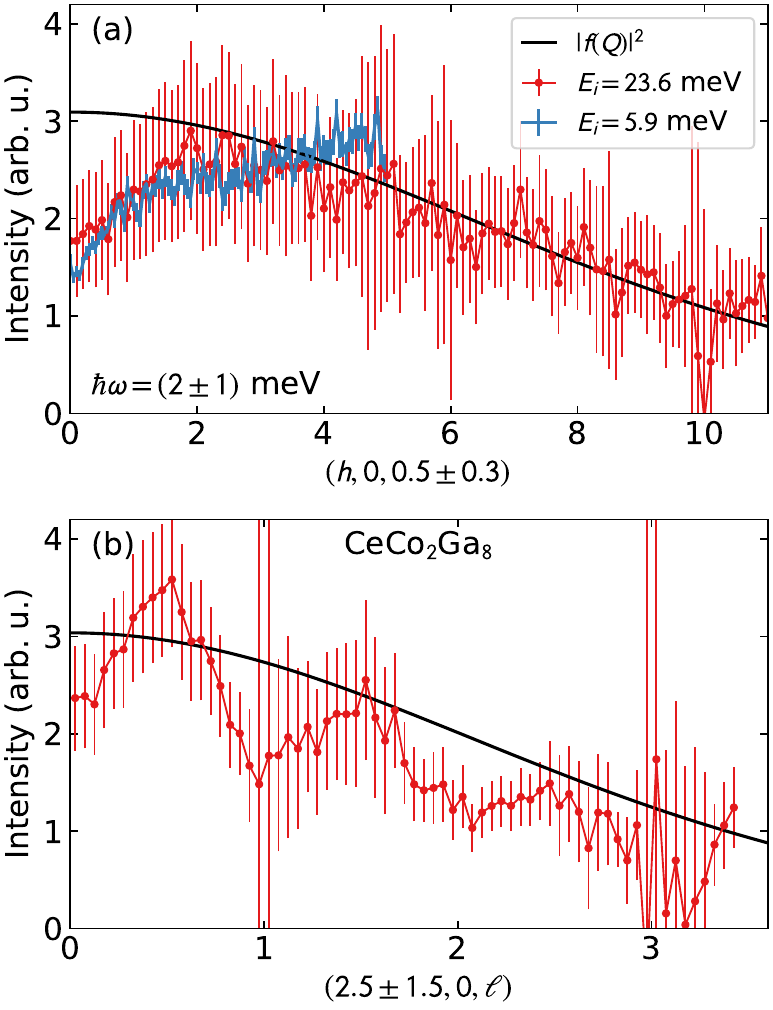}
	\caption{CeCo$_2$Ga$_8$ higher-momentum scattering from $E_i=23.6$~meV data, revealing intensity transverse to the chains that perfectly follows the Ce$^{3+}$ isotropic form factor \cite{BrownFF} in panel (a), indicating no spin correlations between chains. Meanwhile, intensity along the chains in panel (b) follows a sinusoidal modulation commensurate with the lattice, indicating short-ranged antiferromagnetic correlations.}
	\label{fig:FormFactor}
\end{figure}

The CeCo$_2$Ga$_8$ sample measured in neutron spectroscopy is shown in Fig. \ref{fig:crystalmount}. Because of the difficulty in aligning the needle-like samples, they were only aligned along the $c$-direction. Samples were mounted on aluminum plates and held in place with aluminum wire. 
The raw unsymmetrized data are shown in Fig. \ref{fig:unsymmetrized}, while the symmetrized data are shown in Fig. \ref{fig:HOL_comprehensive}. 
Data shown as a function of energy (in e.g. Fig. \ref{fig:unsymmetrized}) show the  $E_i = 2.6$~meV data overlaid over the $E_i = 5.9$~meV data up to 1.5~meV in order to more clearly show the low-energy features. 
Constant energy slices of the  $E_i = 23.6$~meV data are shown in Fig. \ref{fig:Ei23}, which show clearly the modulated intensity along $\ell$ but none along the orthogonal direction. A comparison to the Ce$^{3+}$ form factor is shown in Fig. \ref{fig:FormFactor}, revealing that intensity decays with $|Q|$ precisely as expected for a local Ce doublet, confirming that the inelastic spectrum is indeed magnetic. 

\subsection{Critical scaling}

\begin{figure*}
	\centering
	\includegraphics[width=0.8\textwidth]{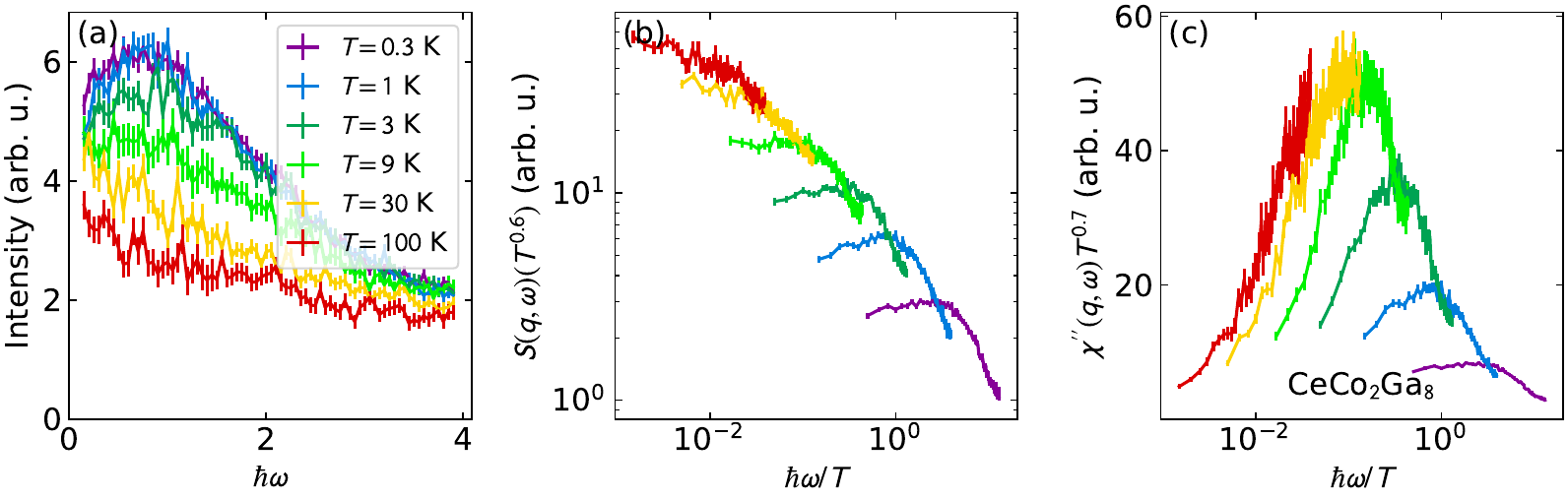}
	\caption{CeCo$_2$Ga$_8$ neutron scattering at $\ell=0.5$ as a function of energy transfer. Panel (a) shows the raw data, panel (b) shows the data scaled by a critical exponent, and panel (c) shows the dynamic susceptibility scaled by a critical exponent. In neither panel (b) nor (c) do the data collapse onto a universal curve, mainly because of the finite-energy maximum in intensity. As shown in Fig. \ref{fig:criticalScaling2}, temperature-subtracting the data does not change this conclusion.}
	\label{fig:criticalScaling1}
\end{figure*}

\begin{figure*}
	\centering
	\includegraphics[width=0.8\textwidth]{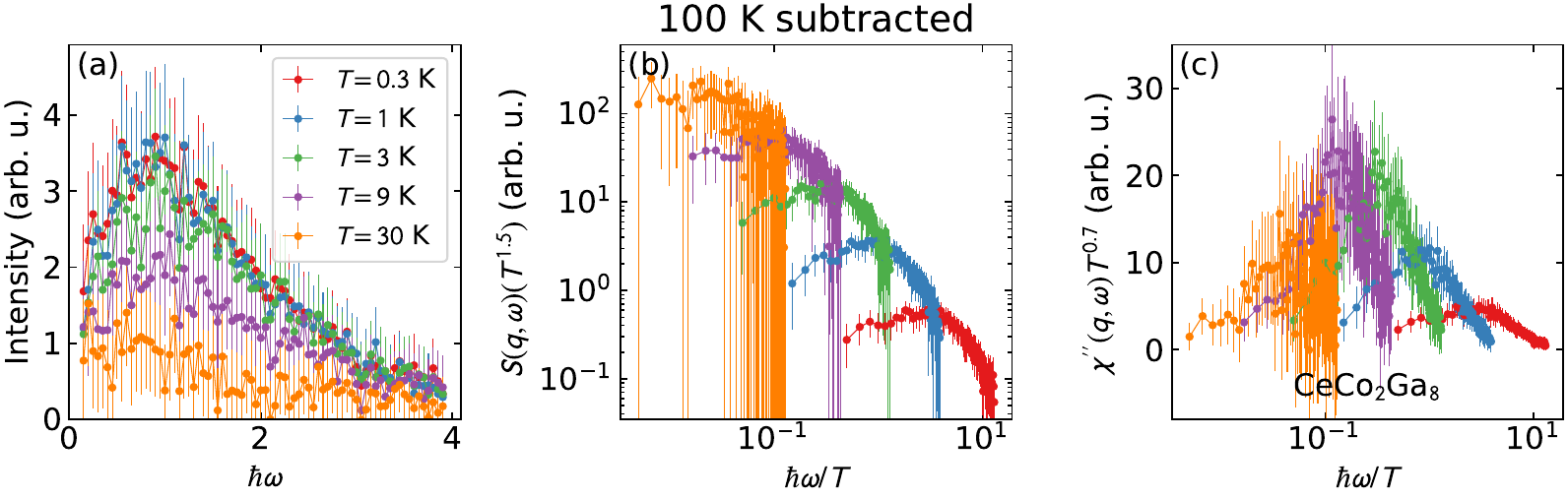}
	\caption{CeCo$_2$Ga$_8$ neutron scattering at $\ell=0.5$ as a function of energy transfer with 100~K data subtracted (c.f. Fig. \ref{fig:criticalScaling1} for the un-subtracted data). In neither panel (b) nor (c) do the data collapse onto a universal curve, mainly because of the finite-energy maximum in intensity.}
	\label{fig:criticalScaling2}
\end{figure*}

To evaluate whether the CeCo$_2$Ga$_8$ scattering exhibits critical scaling, Figs. \ref{fig:criticalScaling1} and \ref{fig:criticalScaling2} show the $\ell=0.5$ scattering from main text Fig. 2 plotted with scaling exponents on the energy and intensity axes. We plot both the neutron structure factor $S({\bf Q},\omega)$ and dynamic susceptibility $ \chi''\left( {\bf Q}, \omega \right)	=
 \pi \left( 1-e^{-\hbar\omega/k_B T} \right) S({\bf Q},\omega) $ where $T$ is temperature and $k_B$ is the Boltzmann constant 
\cite{Lovesey1984}. 
Under no conditions or exponents do the data collapse onto a universal curve, signaling that CeCo$_2$Ga$_8$ is not quantum critical. This is also apparent from the existence of a finite-energy maximum in intensity, which sets a characteristic energy scale to the fluctuations---and thus the fluctuations are not scale-free as in a true quantum critical system. 

\begin{figure}
	\centering
	\includegraphics[width=0.4\textwidth]{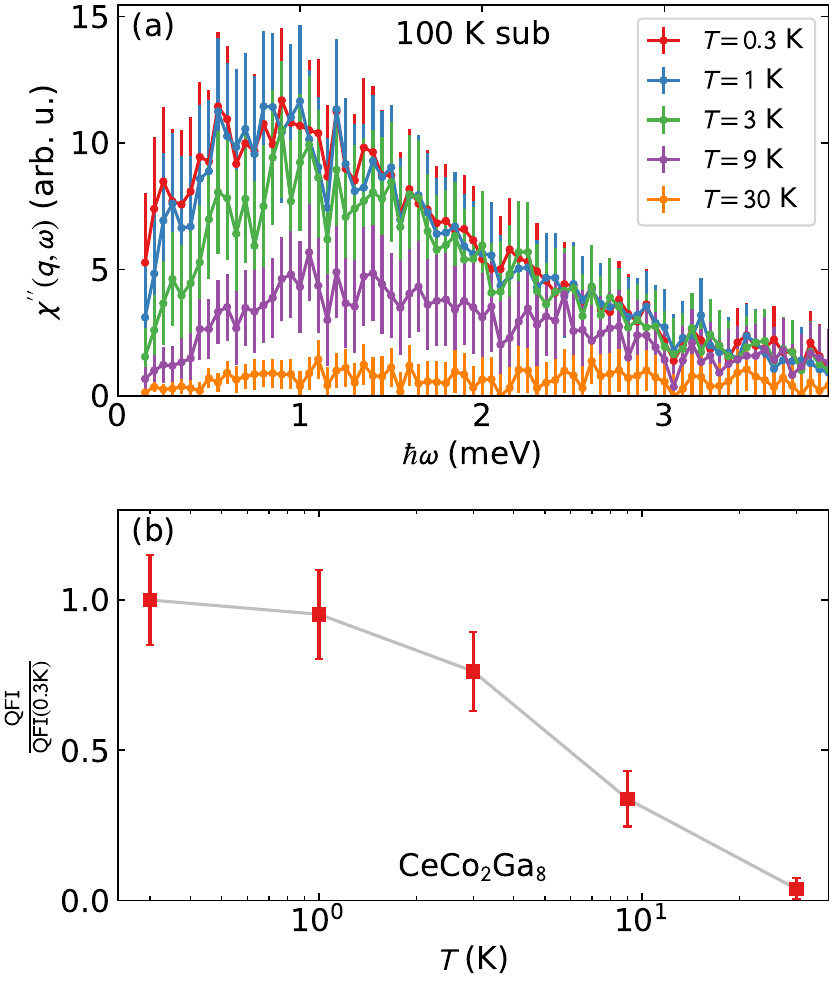}
	\caption{CeCo$_2$Ga$_8$ Quantum Fisher Information (QFI) at $\ell=0.5$. Panel (a) shows the dynamic susceptibility with $T=100$~K data subtracted (as in Fig. \ref{fig:criticalScaling2}). Panel (b) shows QFI calculated with Eq. \eqref{eq:QFI:Hauke}, relative to the lowest $T$ QFI. The data do not show a clear exponential or power law trend over this temperature range.}
    \label{fig:QFI}
\end{figure}

For an alternate perspective on this data, we calculated Quantum Fisher Information (QFI) from the inelastic neutron data \cite{Hauke2016,scheie2025tutorial} using the integral over energy
\begin{equation}
\mathrm{nQFI}\left[ {\bf Q}, T \right]
    = \frac{1}{\pi S^2}	\int_{0}^\infty \mathrm{d}(\hbar \omega) \tanh \left( \frac{\hbar \omega}{2 k_BT }\right) \chi^{\prime\prime} \left( {\bf Q},  \omega\right),	\label{eq:QFI:Hauke}
\end{equation}
where $S$ is the spin quantum number (effectively $S=1/2$ for the Ce$^{3+}$ single-ion ground state doublet). These data are plotted in Fig. \ref{fig:QFI}.  Because the data were not normalized to absolute units, we plot the relative QFI to the $T=0.3$~K data. QFI versus temperature is a way to distinguish quantum phases \cite{shimokawa2025experimentally,zhou2025quantum}, but the QFI versus temperature for CeCo$_2$Ga$_8$ do not follow a simple functional form (exponential or power law). Rather, the QFI seems to saturate at the lowest temperatures. This again indicates that CeCo$_2$Ga$_8$ is either not quantum critical or the spectral weight is not dominantly from critical fluctuations---although the QFI versus $T$ may hold information about the entanglement structure of the many-body quantum state which future theoretical work may elucidate. 

\begin{figure}
	\centering
	\includegraphics[width=0.4\textwidth]{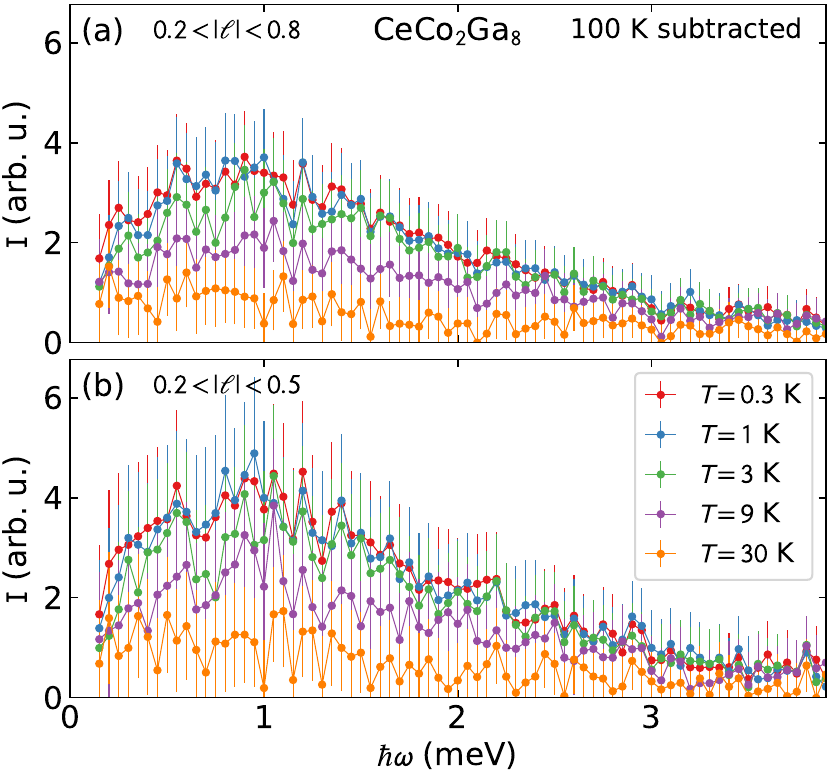}
	\caption{Comparison to energy-dependent scattering in CeCo$_2$Ga$_8$ with different integration ranges along $\ell$. Panel (a) shows $-0.8<\ell<-0.2$ (centered at $\ell=-0.5$) and panel (b) shows $-0.5<\ell<-0.2$ (centered at $\ell=-0.35$, the incommensurate wavevector where intensity is peaked). Besides an overall shift in intensity, there is essentially no difference: both have a peak intensity near 1~meV.}
    \label{fig:CS_integration}
\end{figure}

One may ask whether this analysis changes if we restrict the momentum integration range to be symmetric about the incommensurate wavevector $\ell=0.35$. Figure \ref{fig:CS_integration} shows this is not the case. Restricting the integration range to $-0.5<\ell<-0.2$ still has a clear intensity maximum at 1~meV, precluding any critical scaling. 

\subsection{Fitting the incommensuration}

In the main text Fig. 2, we show that the low-energy scattering shows a split peak indicating incommensurate magnetic correlations. 
The procedure for fitting this was as follows: 
we assume the inelastic fluctuations have a peak shape $C(k,k_0, \alpha, \Gamma)$ that is a numerical convolution of a Gaussian with full-width-half-maximum (FWHM) $\alpha$ and a Lorentzian with FWHM $\Gamma$ centered at $k_0$. Here $\alpha$ represents the instrument resolution and $\Gamma$ gives the correlation length $\xi = \frac{1}{\pi \Gamma}$ (a Lorentzian is a Fourier transform of exponential decay $e^{-\pi \Gamma |x|}$). We also assume that the peaks are centered at $\ell = (n+0.5) \pm \eta$ where $n$ is an integer and $\eta$ is the incommensurate wavevector. 

Because of the noted anisotropy of the spin fluctuations, we must also account for the polarization factor such that $S({\bf q},\omega) = \sum_{\alpha \beta} (\delta_{\alpha \beta} - \hat{q}_{\alpha}\hat{q}_{\beta}) S_{\alpha \beta}$ where $\alpha, \beta \in \{x,y,z \}$ and $\hat{q}_{\alpha} = \frac{q_{\alpha}}{|{\bf q}|}$ is the portion of the unit vector $\bf q$ along $\alpha$ \cite{Squires}. If we neglect off-diagonal spin correlations ($S_{x,y}$, $S_{y,z}$, etc. which are negligible for isotropic exchange), the neutron structure factor is written $S({\bf q},\omega) = \sum_{\alpha} (1 - \hat{q}_{\alpha}^2) S_{\alpha \alpha}$. 
Assuming azimuthal symmetry such that $S_{xx} = S_{yy}$ (where $x,y,z$ are oriented along $h,k,\ell$) and further assuming that the spin fluctuations only change in magnitude along the different directions such that $S_{xx} = B S_{zz}$, we then derive a simplified structure factor with a single adjustable parameter $B$:
\begin{equation}
    S({\bf q},\omega) = \left[2 B(1 - \hat{q}_{x}^2) + (1 - \hat{q}_{z}^2) \right] S_{zz}
\end{equation}
Thus the final fitted equation is 
\begin{equation}
\begin{split}
    f(q)  = & A \left[2 B(1 - \hat{q}_{x}^2) + (1 - \hat{q}_{z}^2) \right] \\
    & \sum_{n} \left[ C(q, n + 0.5 + \eta, \alpha, \Gamma) + C(q, n + 0.5 - \eta, \alpha, \Gamma) \right] 
\end{split}
\end{equation}
which includes the fitted parameters $A$ (overall scale factor), $B$ (spin fluctuation anisotropy), $\eta$ (incommensuration), and $\Gamma$ (inverse of correlation length), plus a constant background offset. 
The fitted values for $\hbar\omega = 0.5$~meV are $\eta = 0.146(2)$~rlu, $B = 0.086(7)$, and $\Gamma = 0.240(7)$~rlu. 
This gives the incommensurate wavevector and correlation length reported in the main text---but the fitted $B$ also gives an estimate of the spin fluctuation anisotropy: over 90\% of the signal is found in the $S_{zz}$ channel, which confirms that the spins are primarily oriented along the $c$ axis. 

\begin{figure}
	\centering
	\includegraphics[width=0.49\textwidth]{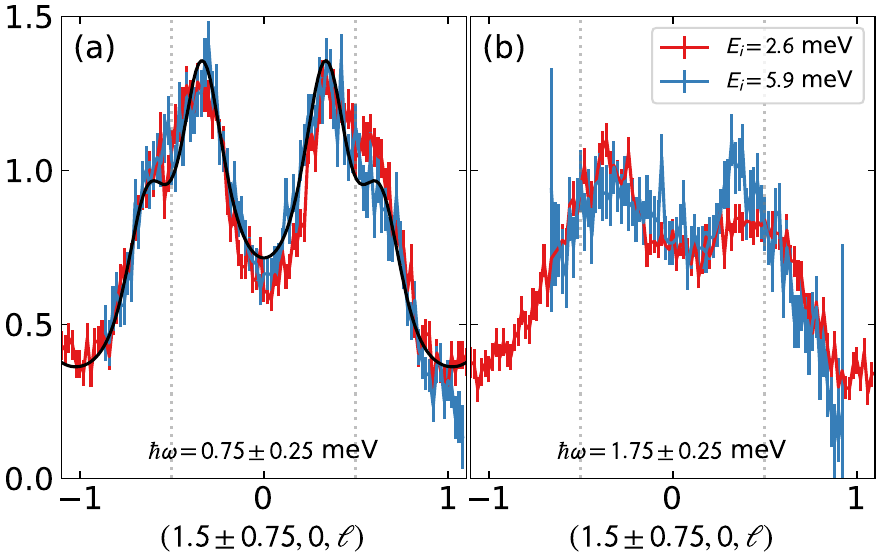}
	\caption{CeCo$_2$Ga$_8$ neutron scattering at low energies showing signs of a split peak around $\ell = 1/2$. Panel (a) shows the scattering at $\hbar \omega = 0.75$~meV, while panel (b) shows the scattering at  $\hbar \omega = 1.75$~meV. The black line is the best fit from the main text, and the grey dashed lines show the commensurate $\ell=\pm1/2$. Despite the slightly different integration ranges here versus the main text, the fit still captures the features amd the incommensurate peak intensity.}	
    \label{fig:incommensuration}
\end{figure}

\begin{figure*}
	\centering
	\includegraphics[width=0.79\textwidth]{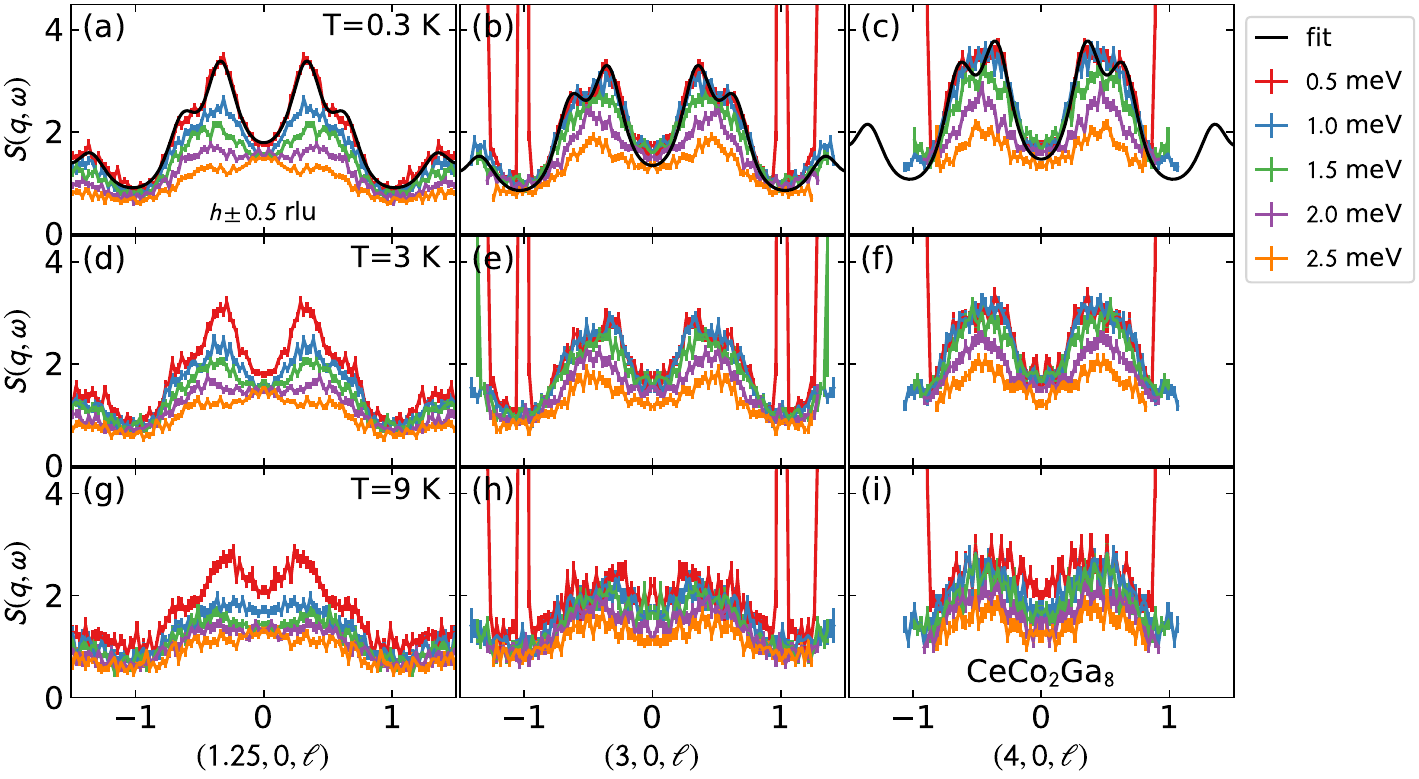}
	\caption{CeCo$_2$Ga$_8$ $\ell$-dependent neutron scattering at various temperatures and energies. Panels  (a)-(c) show 0.3~K, panels (d)-(f) show 3~K, and panels (g)-(i) show 9~K. The black line is the best fit from the main text, form-factor adjusted to the various values of $h$. As temperature increases, the incommensuration survives, but the peaks broaden and become less intense.}	
    \label{fig:incommensuration2}
\end{figure*}

To demonstrate that this split peak is not an artifact of energy choice, Fig. \ref{fig:incommensuration} shows the $\ell$-dependent scattering at different incident energies. In all cases, the peak intensity is not found at $\ell=1/2$, but at slightly incommensurate wavevectors. This is also evident from the steepness of sides of the intensity feature, which is definitely not captured by a Gaussian or Lorentzian at $\ell=1/2$. The temperature dependence of these features is shown in Fig. \ref{fig:incommensuration2}, revealing the incommensuration survives to appreciable temperatures.  

\section{ARPES details}

\begin{figure}
	\centering
	\includegraphics[width=\columnwidth]{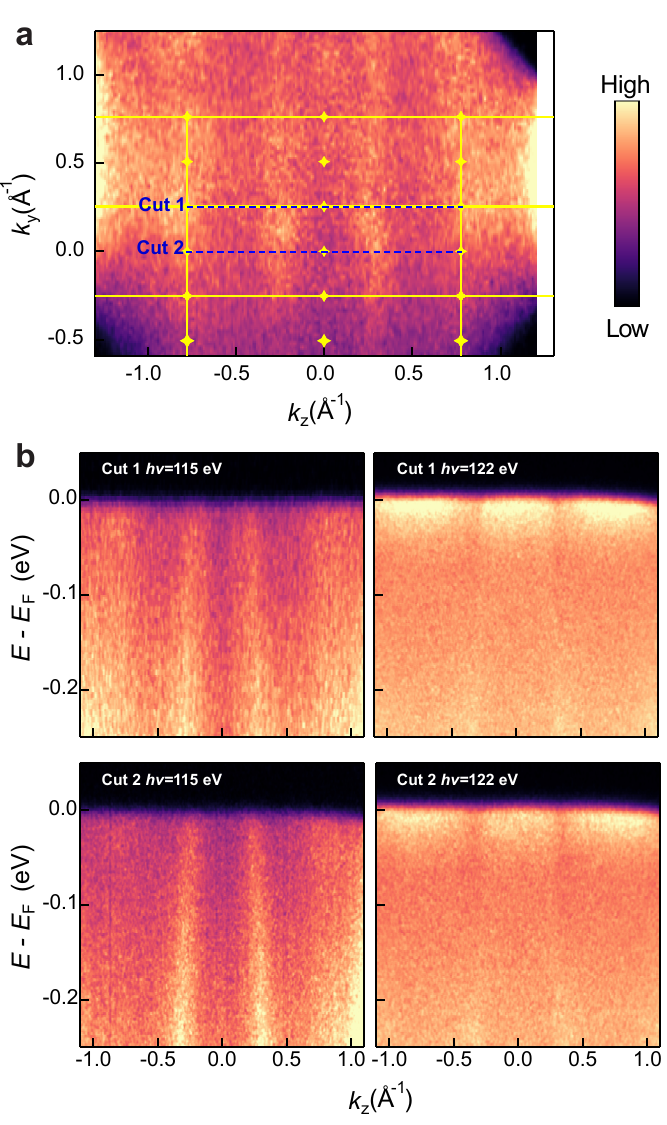}
	\caption{Transverse electronic dispersion of CeCo$_2$Ga$_8$. 
    {\bf a} In-plane $k_z–k_y$ Fermi surface of a second CeCo$_2$Ga$_8$ sample measured at $h\nu = 115$~eV. Weakly modulated Fermi sheets are observed near the zone center. The Brillouin zone is indicated in yellow. The Fermi surface was acquired at $T = 8$~K and integrated within $\pm 15$~meV of $E_F$. \
    {\bf b} Energy-momentum dispersions measured along the cuts indicated in {\bf a}, using photon energies of $h\nu = 115$~eV and 122~eV. Hole-like dispersive states are observed near $kz = \pm 0.25$~\AA$^{-1}$. The dispersions and Fermi-crossing positions show minimal variation between different in-plane momentum cuts and between the two photon energies, indicating weak dispersion along both $k_x$ and $k_y$. All data were measured using linear vertical polarization.}
	\label{fig:ARPES-SI}
\end{figure}

To further examine the transverse dispersion of the low-energy electronic states in ARPES, we acquired energy-momentum cuts at several $k_y$ positions on a second CeCo$_2$Ga$_8$ sample, as indicated in SI Fig. \ref{fig:ARPES-SI}(a). Representative dispersions measured at photon energies of $h\nu = 115$~eV and 122~eV are shown in SI Fig. \ref{fig:ARPES-SI}(b). In each case, hole-like dispersive states are observed near $k_z = \pm0.25$~\AA$^{-1}$.     Strong $f$-weight near the Fermi level can be observed by the on-resonance data obtained at $h \nu = 122$~meV, reflecting the presence of Kondo scattering. The dispersion, spectral weight distribution, and Fermi-crossing positions remain nearly unchanged between different in-plane momentum cuts. This weak dependence on $k_y$ indicates strongly suppressed in-plane transverse dispersion. Similarly, the absence of significant changes between the two photon energies indicates weak $k_x$  dispersion. These measurements therefore support the conclusion that the low-energy electronic structure of CeCo$_2$Ga$_8$ is highly anisotropic, with strongly reduced dispersion along both transverse momentum directions. 

We fit momentum distribution curves (MDCs) extracted from the $E-k$ dispersion shown in SI Fig.~\ref{fig:ARPES-SI}(b) (Cut 2, h$\nu$~=~115~eV) to determine the Fermi wave vector ($k_{\mathrm F}$) and Fermi velocity ($v_{\mathrm F}$). These quantities were mapped onto a one-dimensional cosine dispersion, $E(k)=E_0+t\cos(ka)$, for which
$|\hbar v_{\mathrm F}|=ta|\sin(k_{\mathrm F}a)|$.

Extrapolating the fitted dispersions to $E_{\mathrm F}=0$~eV and averaging the positive- and negative-momentum branches gives $\langle |k_{\mathrm F}| \rangle = 0.234\pm0.001~\mathrm{\AA^{-1}}$ and $
\langle |\hbar v_{\mathrm F}| \rangle = 1.8\pm0.2~\mathrm{eV,\AA}$. Using $a=4.059~\mathrm{\AA}$, these values yield an effective hopping parameter $\langle |t| \rangle = 0.56\pm0.07~\mathrm{eV}$. The ARPES-derived hopping parameter is therefore consistent with the DFT value of t=0.5~eV. 

\section{DMRG simulations}

We use $U(1)$ DMRG methods~\cite{white1992density,white1993density,schollwock2011density} to simulate the ground state and spectrum of the 1D Kondo lattice model. Calculations are performed using the ITensor library v3~\cite{itensor}. In practice we use $L=48$ and retain up to $M=600$ bond dimensions for the ground state calculations to reach a typical truncation error in the order of $10^{-6}$. The time evolution is done by the one-site TDVP methods~\cite{haegeman2011time,haegeman2016unifying} with enlarged bond dimension through global Krylov vectors~\cite{yang2020time}. Because the entanglement entropy of the state grows with time evolution, the total time that can be faithfully accessed during the simulations is limited by the bond dimensions of the state. For smaller $J_{H}$ or $J_k$ the energy scale of the spectral function is typically smaller, and $\delta \omega = 2\pi / T$ is bounded by the total time in the discrete Fourier transformation. Thus, longer time is needed to obtain the spectral function. At $J_{H}=0.05t$ bond dimensions up to $M=1600$ are used to obtain simulation time up to $T=450$ in the unit of $1/t$. We measure the correlation function for every time step of $\delta \tau =0.5/t$.

\begin{figure*}
	\centering
	\includegraphics[width=0.8\textwidth]{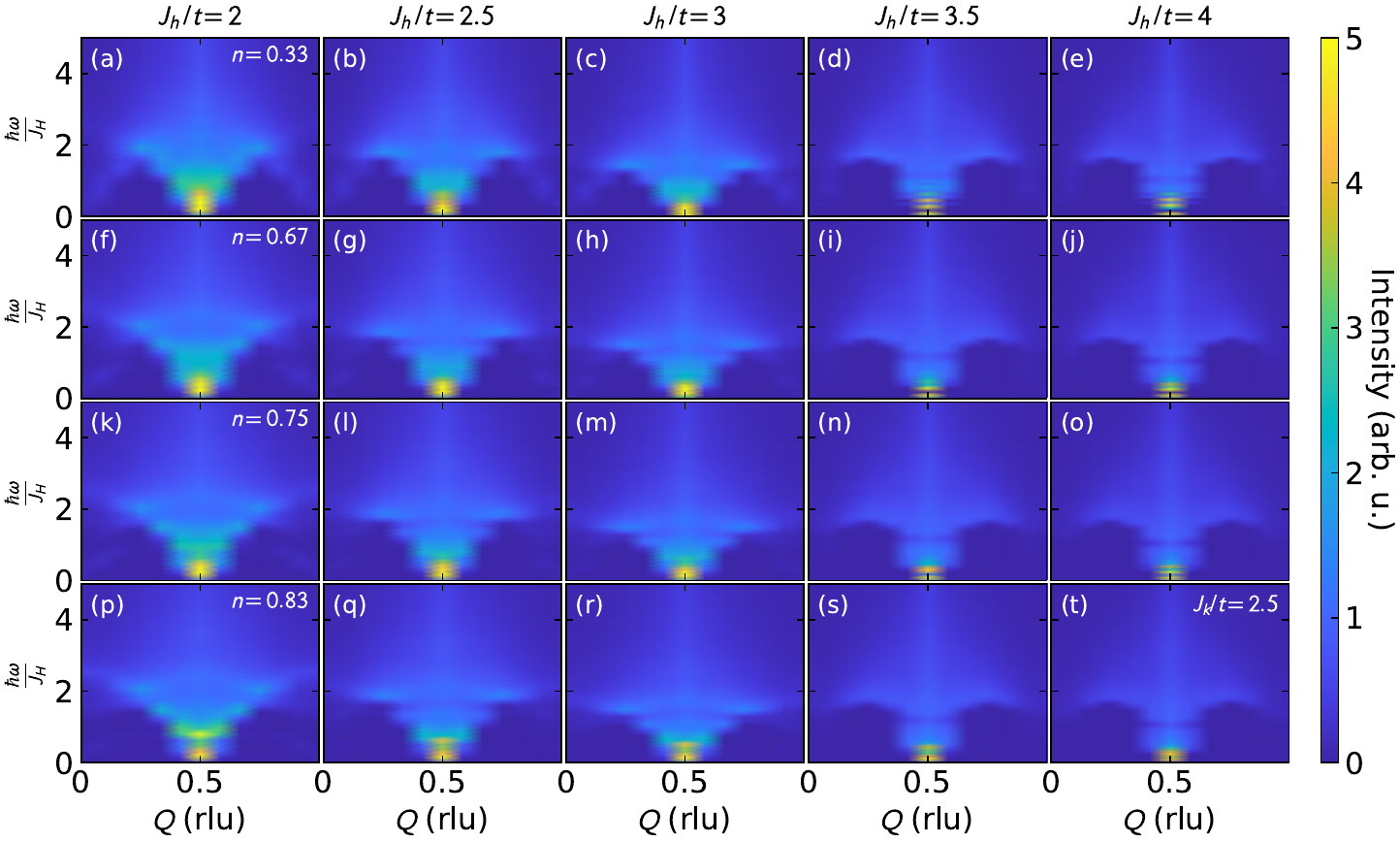}
	\caption{DMRG simulations of the Kondo lattice model with various $J_H$ and $n$ values. All these simulations lie within the 1D AFM ("spin gap") phase, and to varying degrees resemble the 1D Heisenberg chain spectrum \cite{caux2006four}. Although the spectra vary slightly with different values of $n$, it does not make a dramatic difference to the calculated spectra.}
	\label{fig:DMRG1}
\end{figure*}

\begin{figure*}
	\centering
	\includegraphics[width=0.99\textwidth]{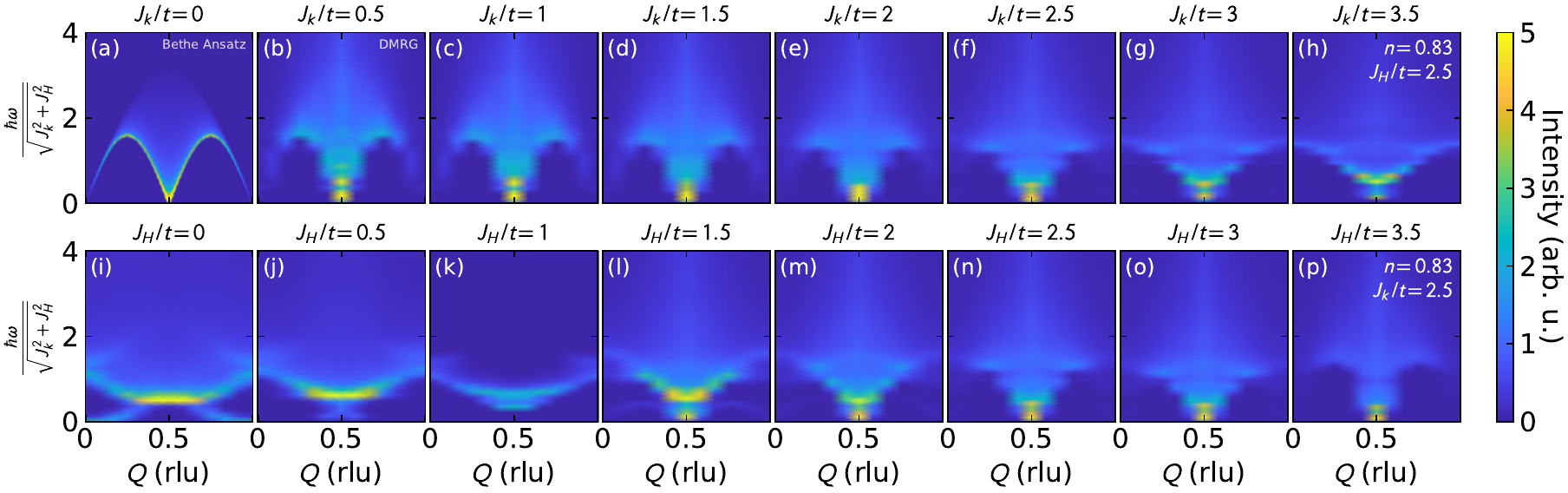}
	\caption{DMRG simulations of the Kondo lattice model tuning the $J_H$ and $J_K$ values. The top panel (a)-(h) is completely within the 1D AFM ("spin gap") phase in Ref. \cite{Sikkema_1997}, and includes the Bethe Ansatz solution for $J_K=0$ \cite{caux2006four}, where the system is the ideal 1D Heisenberg chain. The system tunes from a linear dispersive spinon spectrum to a linear dispersive spectrum gapped at $Q=0$. The bottom panel (i)-(p) is also plotted in main text Fig. 3, and shows tuning from the Luttinger Liquid to the 1D AFM ("spin gap") phase. }
	\label{fig:DMRG2}
\end{figure*}

Although the Friedel oscillation is expected in the ground state of the Luttinger liquid phase~\cite{shibata1996friedel} which induces a finite $<S_{i}^{z}>$ at the open boundary that decays into the center of the chain, we find that the momentum-frequency resolved spectral function has little boundary effect. The spectral function is almost the same with $L_{0}=L/2$ which is closer to the open boundary than with $L_{0}=L/4$.
Some representative spectral functions are shown in Figs. \ref{fig:DMRG1} and \ref{fig:DMRG2}.

\begin{figure}
\centering
\includegraphics[width=\linewidth]{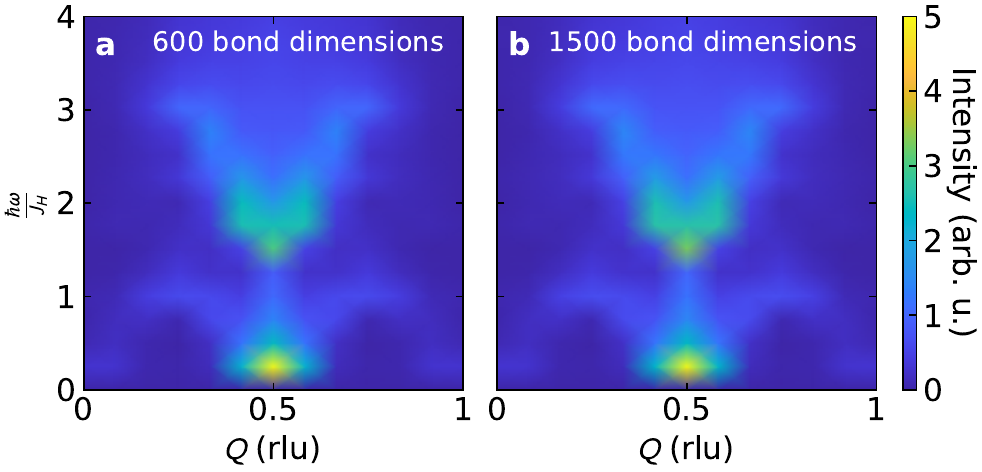}
\caption{DMRG calculated spectral functions using various bond dimensions at $J_{K}=1.6$, $J_{H}=0.5$ and $n=5/6$ on a $L=48$ lattice.}
\label{Figs_convergence}
\end{figure}

To ensure numerical convergence, we test the spectral functions with different bond dimensions as shown in Fig.~\ref{Figs_convergence}. The results are almost the same for 600 and 1500 bond dimensions at $J_{K}=1.6$, $J_{H}=0.5$. For smaller $J_{H}=0.05$ we obtain the results with larger bond dimensions up to 1600 in order to simulate long time evolutions.

We also test more realistic values of $J_{H}$ that are two or three orders of magnitude smaller than $t$. In the possible incommensurate regime, the DMRG optimization primarily updates the electron degrees of freedom during the sweeps while the localized spins become trapped in a local minimum. By contrast, the 1D AFM phase converges reliably even for $J_{H}\approx 0.01t$. However, convergence becomes much worse when $J_{H}$ is tuned down into the incommensurate phase. A similar convergence issue is found in other tensor-network-based methods, such as infinite DMRG and variational uniform matrix product state (VUMPS) algorithm. 

\begin{figure}
\centering
\includegraphics[width=\linewidth]{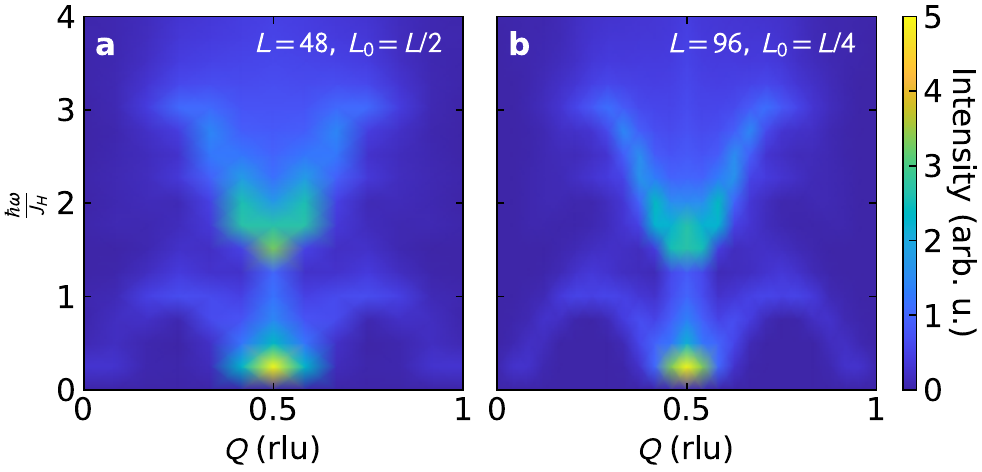}
\caption{DMRG calculated spectral functions at $J_{K}=1.6$, $J_{H}=0.5$ and $n=5/6$ on the lattice of (a) $L=48, L_{0}=L/2$ and (b) $L=96, L_{0}=L/4$.}
\label{Figs_size}
\end{figure}

To avoid the open boundary effect, we choose a segment of $L_{0}=L/4$ in the middle of the chain to calculate the momentum-resolved spectral function. A similar results can be obtained with $L_{0}=L/2$, as shown in Fig.~\ref{Figs_size} (a). In addition, we also calculate the spectral functions on a larger lattice with $L=96$ where the dispersion becomes more smooth in the high energies. Overall, the dispersion of the spectral functions remains almost the same for different lattice sizes, suggesting very little finite size effect.

\begin{figure*}
\centering
\includegraphics[width=0.75\linewidth]{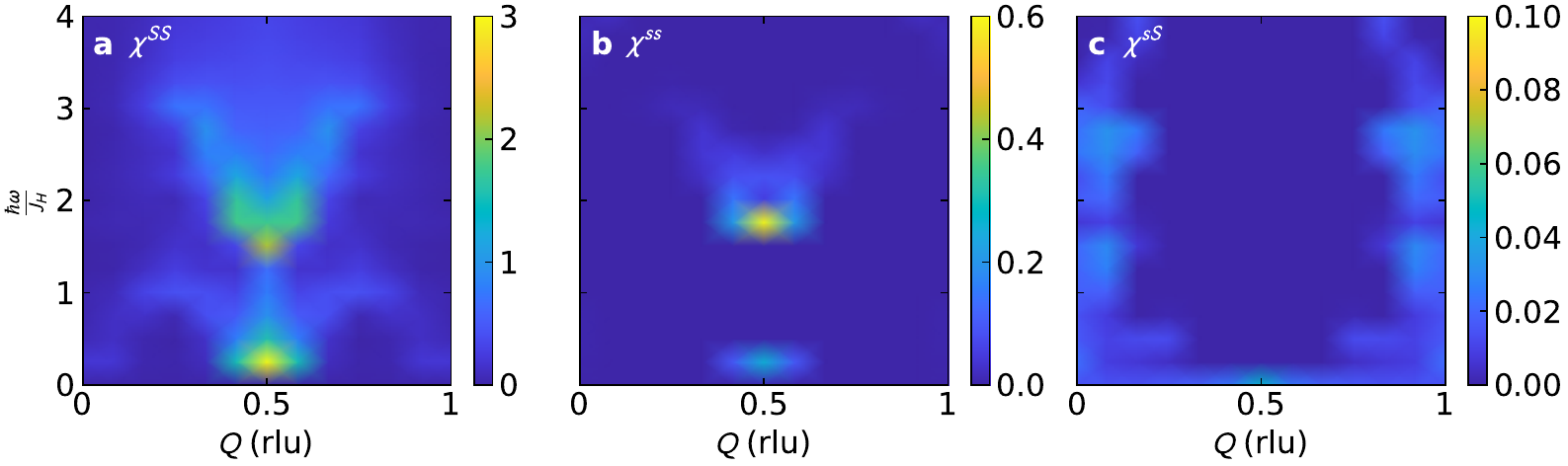}
\caption{DMRG calculated spectral functions of (a) localized spins, (b) electron spins, and (c) the mixed electron spin-localized spin contributions at $J_{K}=1.6$, $J_{H}=0.5$ and $n=5/6$ on a $L=48$ lattice. Note the different scales on the colorbars, and that the $\chi^{ss}$ is an order of magnitude weaker than $\chi^{SS}$. and the $\chi^{sS}$ is weaker still.}
\label{Figs_compare_16}
\end{figure*}

\begin{figure*}
\centering
\includegraphics[width=0.75\linewidth]{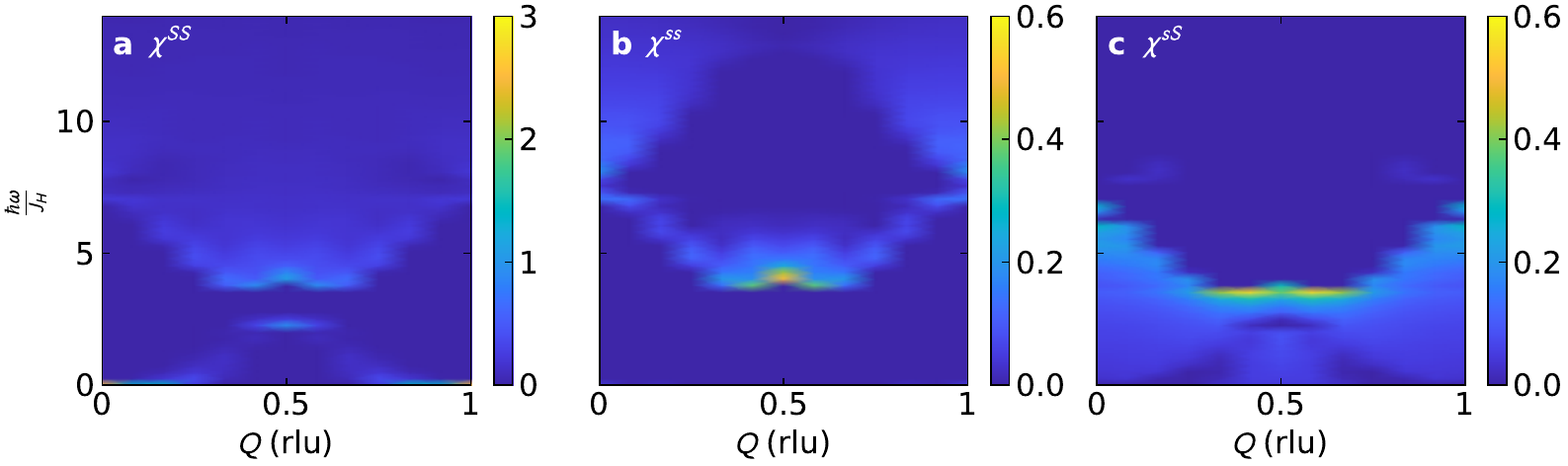}
\caption{DMRG calculated spectral functions of (a) localized spins, (b) electron spins, and (c) the mixed electron spin-localized spin contributions at $J_{K}=3$, $J_{H}=0.5$ and $n=5/6$ on a $L=48$ lattice. Note the different scales on the colorbars. In this case the $\chi^{sS}$ is comparable to $\chi^{ss}$, but still much weaker than $\chi^{SS}$ (which has a strong intensity at $\hbar \omega = 0$ and $Q=0$). }
\label{Figs_compare_3}
\end{figure*}

In the main text we show the spectral functions of the localized spin $\chi^{SS}(q,\omega)$ which always dominates the electron spin-spin $\chi^{ss}(q,\omega)$ and the mixed electron spin-localized spin $\chi^{sS}(q,\omega)$ in the low energy spectrum. 
In practice we find that the $\chi^{SS}(q,\omega)$ always dominates the low energy spectrum in both the Luttinger liquid phase and the 1D spin chain limit. 
As shown in Fig.~\ref{Figs_compare_16}, the spectral functions of the localized spins $\chi^{SS}(q,\omega)$ at low energies is at least one order of magnitude larger than $\chi^{ss}(q,\omega)$ and $\chi^{sS}(q,\omega)$ in the 1D spin chain limit. In addition, we show in Fig.~\ref{Figs_compare_3} that the $\chi^{SS}(q,\omega)$ at low energies dominates $\chi^{ss}(q,\omega)$ and $\chi^{sS}(q,\omega)$ in the Luttinger liquid phase by more than one order of magnitude.
This can be expected as the 1D spin chain limit is dominated by the Heisenberg exchange of the localized spins and the Luttinger liquid phase is dominated by the effective RKKY interactions~\cite{ruderman1954indirect,kasuya1956theory,yosida1957magnetic} of the localized spins. Thus in the main text plots we neglect the electron spin-spin correlations and mixed electron spin-localized spin correlations.

\end{document}